\documentclass[12pt]{report}
\usepackage{latexsym,amsmath,amssymb,color}
\usepackage{ifthen}

\def\F{I\kern-.30em{F}}
\def\P{I\kern-.30em{P}}
\def\E{I\kern-.30em{E}}
\def\build#1_#2^#3{\mathrel{\mathop{\kern 0pt#1}\limits_{#2}^{#3}}}
\def\Sum{\displaystyle\sum}

\def\supp{\mbox{\rm supp}\ }

\def\dsp{\displaystyle}
\def\beq{\begin{equation}}
\def\eeq{\end{equation}}
\def\ba{\begin{array}}
\def\ea{\end{array}}
\def\boa{\left.\begin{array}{llll}}
\def\eoa{\end{array}\right.}
\def\bc{\begin{center}}
\def\ec{\end{center}}

\def\qed{\ {\vrule height5pt depth0pt width5pt}\ } 

\catcode`\@=11
\newif\ifproofmode
\proofmodetrue
\def\labelcour{}

\def\@endtheorem{\hfill\ifproofmode\rlap{\tiny \kern5mm
\labelcour}\fi\endtrivlist}

\catcode`\@=12

\newcommand{\Cm}[2][0]{\ensuremath{C^{#1}(#2)}}

\newcommand{\demi}{\ensuremath{\frac{1}{2}}}
\newcommand{\demip}{\ensuremath{1 \slash 2}}

\newcommand{\quartp}{\ensuremath{1 \slash 4}}

\newcommand{\expo}[1]{\ensuremath{{\rm e}^{#1}}}

\newcommand{\Lp}[2][2]{\ensuremath{L^{#1}(#2)}}
\newcommand{\Lps}[3][2]{\ifthenelse{\equal{#2}{0}}
                                   {\ensuremath{L^{#1}(#3)}}
                                   {\ensuremath{L^{#1}_{#2}(#3)}}}
\newcommand{\Lploc}[2][2]{\ensuremath{L_{loc}^{#1}(#2)}}
\newcommand{\Hploc}[2][1]{\ensuremath{H_{loc}^{#1}(#2)}}
\newcommand{\Hp}[2][1]{\ensuremath{H^{#1}(#2)}}

\newcommand{\R}{\mathbb{R}}

\newcommand{\Z}{\mathbb{Z}}

\newcommand{\N}{\mathbb{N}}

\newcommand{\C}{\mathbb{C}}

\newtheorem{theorem}{Theorem}[section]
\newtheorem{lemma}{Lemma}[section]
\newtheorem{proposition}{Proposition}[section]
\newtheorem{corollary}{Corollary}[section]

\newtheorem{lem:peter}{Lemma}

\newcommand{\Schr}{Schr{\"o}dinger}

\newcommand{\bea}{\begin{eqnarray}}
\newcommand{\eea}{\end{eqnarray}}

\begin{document}

\begin{titlepage}

\vspace*{-1.0in}
\begin{center}

{\bf Edge Currents for Quantum Hall Systems, \\
I.\ One-Edge, Unbounded Geometries }

  \vspace{0.3 cm}

  \setcounter{footnote}{0}
  \renewcommand{\thefootnote}{\arabic{footnote}}

  {\bf Peter D.\ Hislop \footnote{Supported in part by NSF grant
      DMS-0503784.}}

  \vspace{0.1 cm}

  {Department of Mathematics \\
    University of Kentucky \\
    Lexington, KY 40506--0027 USA}

  \vspace{0.1 cm}

  {\bf Eric Soccorsi \footnote{also Centre de Physique Th\'eorique, Unit\'e
Mixte de Recherche 6207 du CNRS et des Universit\'es Aix-Marseille I,
Aix-Marseille II et de l'Universit\'e du Sud Toulon-Var-Laboratoire
affili\'e \`a la FRUMAM, F-13288 Marseille Cedex 9, France.}}

  \vspace{0.1 cm}

  {Universit\'e de la M\'editerran\'ee \\
      Luminy, Case 907 \\
  13288 Marseille, FRANCE}

\end{center}

\vspace{0.1 cm}

\begin{center}
  {\bf Abstract}
\end{center}

\noindent
Devices exhibiting the integer quantum Hall effect can be modeled
by one-electron Schr\"odinger operators describing the planar motion of an
electron in a perpendicular, constant magnetic field, and under the
influence of an electrostatic
potential. The electron motion is confined to unbounded subsets of the
plane by confining potential barriers. The edges of the confining potential
barrier create edge currents.
In this, the first of two papers, we
prove explicit lower
bounds on the edge currents associated with one-edge, unbounded
geometries formed by various confining
potentials. This work extends some known results that we review.
The edge currents
are carried by states with energy localized between any two Landau
levels. These one-edge geometries
describe the electron confined to certain unbounded regions
in the plane obtained by deforming half-plane regions.
We prove that the currents are stable under various
potential perturbations, provided the perturbations
are suitably small relative to
the magnetic field strength, including perturbations by random potentials.
For these cases of one-edge geometries,
the existence of, and
the estimates on, the edge currents imply that the corresponding
Hamiltonian has intervals of absolutely continuous spectrum.
In the second paper of this series, we consider the edge currents
associated with two-edge geometries
describing bounded, cylinder-like
regions, and unbounded, strip-like, regions.

\vspace{0.2 cm}

\end{titlepage}


\tableofcontents


\renewcommand{\thechapter}{\arabic{chapter}}
\renewcommand{\thesection}{\thechapter}

\setcounter{chapter}{1} \setcounter{equation}{0}

\section{Introduction and Main Results}

The integer quantum Hall effect (IQHE) refers
to the quantization of the Hall conductivity
in integer multiples of $2 \pi e^2 / h$.
The IQHE is observed in planar quantum devices at zero temperature
and can be described by a Fermi gas of noninteracting electrons.
This simplification reduces the study of the dynamics to
the one-electron approximation.
Typically, experimental devices consist of finitely-extended,
planar samples subject to a constant perpendicular
magnetic field $B$.  An applied electric field in the $x$-direction induces
a current in the $y$-direction, the Hall current,
and the Hall conductivity $\sigma_{xy}$ is observed to be quantized.
Furthermore, the Hall conductivity is a
function of the electron Fermi energy, or, equivalently, the
electron filling factor, and plateaus of the Hall conductivity are
observed as the filling factor is increased.
It is now accepted that the
occurrence of the plateaus is due to the existence of localized
states near the Landau levels that are created by the
random distribution of impurities
in the sample, cf.\ \cite{[BESB1]}.

Another new phenomenon that arises in the
study of these devices exhibiting the IQHE is the
occurrence of {\it edge currents} associated with the boundaries of
quantum devices.  These edge currents are the subject of this work.
In order to explain their origin, we recall the theory of an electron in
$\R^2$
subject to a constant, transverse magnetic field.
The Landau Hamiltonian $H_L(B)$ describes a charged
particle constrained to $\R^2$,
and moving in a constant, transverse magnetic field with strength $B \geq
0$. Let $p_x = -i \partial_x$ and $p_y = - i \partial_y$ be the two
momentum operators. The operator $H_L(B)$
is defined on the dense domain $C_0^\infty (\R^2) \subset L^2 ( \R^2)$ by
\beq
\label{defnop1}
H_L(B) = (-i \nabla - A)^2 = p_x^2 + (p_y - Bx)^2 ,
\eeq
in the Landau gauge for which the vector potential is $A(x,y) = B(0, x)$.
The map (\ref{defnop1})
extends to a self-adjoint operator with point spectrum given by $\{
E_n (B)  = (2n + 1)B \; | \; n =0 , 1, 2, \ldots \}$, called the {\it
Landau levels}, and each eigenvalue is infinitely degenerate.
The perturbation of $H_L (B)$ by random Anderson-type potentials $V_\omega$
in the weak disorder regime for which $\| V_\omega \| < C_0 B$
has been extensively studied, cf.\
\cite{[CH2],[DMP],[GK1],[Wang1]}.
It is proved that outside a small interval of size $B / \log B$
about the Landau levels,
there are intervals of
pure point
spectrum with exponentially decaying eigenfunctions. The nature of the
spectrum at the Landau levels is unclear. It is now known that there is
nontrivial transport near the Landau levels for models on $L^2 (\R^2)$
\cite{[GKS]}.
For a point interaction model on the lattice $\Z^2$, studied in
\cite{[DMP2]},
the authors considered the first $N$ Landau levels and proved that there
exists an $B_N > 0$ so that if $B > B_N$, then the spectrum of $H_\omega$
below the $N^{th}$ Landau level is pure point almost surely
and that
each Landau level below the $N^{th}$ is infinitely degenerate.

The quantum devices studied
with regard to the IQHE may be infinitely extended or finite, but
are distinguished by the fact that there is at least one edge,
that can be considered infinitely extended, like in the case of the
half-plane, or periodic, as in case of an annulus
or cylinder.
In all cases, the unperturbed Hamiltonian is a nonnegative, self-adjoint
operator on the Hilbert space $L^2 (\R^2)$ and having the form
\beq
\label{unpert1}
H_0 = H_L (B) + V_0 ,
\eeq
where $V_0$ denotes the confining potential forming the edge (we also
consider Dirichlet boundary conditions).
The existence of an edge
profoundly changes the transport and spectral properties
of the quantum system.
We consider states $\psi \in L^2 (\R^2)$ with energy concentration between
two successive Landau levels $E_n(B)$ and $E_{n+1} (B)$.
We say that such a state $\psi$ carries an
{\it edge current} if the expectation of the $y$-component of
the velocity operator $V_y \equiv (p_y - Bx)$ in the state $\psi$ is
nonvanishing.
In these two papers, we prove the existence of edge currents carried by
these states and provide an explicit lower bound on the strength of the
current. This lower bound shows that the edge current persists for all
time in that the expectation of the Heisenberg time-dependent current
operator $V_y(t) \equiv e^{itH} V_y e^{-itH}$ in the state $\psi$ satisfies
the same lower bound for all time.
We will also prove that the states that carry edge-currents are
well-localized in a neighborhood of the boundary of the region.

Our main results, presented in this paper and its sequel,
concern the following geometries and confining potentials.

\begin{enumerate}
\item One-Edge Geometries: We study the half-plane case for which the
electron is constrained to the right half-plane $x > 0$ by a confining
potential $V_0$ that has either of the two forms:
\begin{enumerate}
\item Hard Confining Potentials, such as the Sharp Confining Potential:
  $V_0(x) = \mathcal{V}_0 \chi_{\{x < 0\}} (x) $, where $\mathcal{V}_0 > 0$
is a constant, or Dirichlet boundary
  conditions along the edge $x = 0$.
\item Soft Confining Potentials, such as the Parabolic Confining Potential:
$V_0 (x) = \mathcal{V}_0 x^2  \chi_{\{x < 0\}} (x)$, and other rapidly
increasing confining potentials.
\end{enumerate}
\item Two-Edge Geometries: We study models for which the electron is
confined to the strip $B_L = [ -L/2 , L/2] \times
\R$ by hard or soft confining potentials, such as
\begin{enumerate}
\item Sharp Confining Potential: $V_0(x) = \mathcal{V}_0
   \chi_{\{ |x| > L/2 \} } (x) .$
\item Polynomial Confining Potential:  $V_0 (x) = \mathcal{V}_0 ( |x| -
   L/2 )^p \chi_{\{ |x| > L/2 \} } (x)$, for $p > 1$.
\end{enumerate}
\item Bounded, Two-Edge Geometries: We study
models that are topologically a cylinder
$\R \times S^1$ with confining potentials along the $x$-direction.
\end{enumerate}
The present paper deals with the first topic of one-edge geometries,
and the sequel \cite{[HS2]} deals with the second and third topics
concerning
two-edge geometries.

In addition to these results for straight edge geometries,
we show that the results are stable under certain
perturbations of the straight edge boundaries.
Concerning the hard confining potentials,
we note that the lower bounds for the Sharp Confining Potential
are uniform with respect to the
strength of the confining potential $\mathcal{V}_0$.
This means that we can take the limit as the size of
the confining potential becomes infinite. As a result, our results extend
to the case of Dirichlet boundary conditions along the edges.
The various soft confining potentials are discussed in section 6.

Our strategy in the one-edge case
is to analyze the unperturbed operator via the partial Fourier
transform in the $y$-variable. We write $\hat{f} (x, k)$ for this
partial Fourier transform.
This decomposition reduces the problem to a study of the
fibered operators of the form
\beq
\label{unpert2}
h_0(k) = p_x^2 + (k - Bx)^2 + V_0 (x) ,
\eeq
acting on $L^2(\R)$.
Since the effective,
nonnegative, potential $V (x;k) = (k - Bx)^2 + V_0 (x)$ is unbounded
as $|x| \rightarrow \infty$, the resolvent of $h_0(k)$ is compact and the
spectrum is discrete.
We denote the eigenvalues of $h_0(k)$
by $\omega_j (k)$, with corresponding normalized
eigenfunctions $\varphi_j (x ; k)$, so that
\beq
\label{ev1}
h_0(k) \varphi_j (x;k) = \omega_j(k) \varphi_j (x;k) , ~~~\| \varphi_j (
\cdot; k) \|
= 1.
\eeq

The properties of the eigenvalue maps
$k \in \R \rightarrow \omega_j (k)$ play an
important role in the proofs. These maps are called the {\it dispersion
curves} for the unperturbed Hamiltonian (\ref{unpert1}).
The importance of the properties of the dispersion curves comes from an
application of the Feynman-Hellmann formula. To illustrate this, let us
consider the one-edge geometry of a half-plane
with a sharp confining potential that is treated in this
paper.
It is clear from the form of the effective potential $V (x;k)$
that the dispersion curves are monotone decreasing functions of
$k$, and that
$\lim_{k \rightarrow + \infty} \omega_n (k) = E_n (B)$, and that $\lim_{k
\rightarrow - \infty}\omega_n (k) = E_n(B) + \mathcal{V}_0$.
For simplicity, we consider in this introduction
a closed interval $\Delta_0 \subset (B , 3B)$ and a normalized
wave function $\psi$ satisfying $\psi = E_0 ( \Delta_0 ) \psi$. Such a
function
admits a decomposition of the form
\beq
\label{fourier1}
\psi (x, y) =  \frac{1}{\sqrt{2 \pi}} \int_{\omega_0^{-1} (\Delta_0)}
e^{i k y} \beta_0 (k) \varphi_0 (x ; k) ~dk,
\eeq
where the coefficient $\beta_0 (k)$ is defined by
\beq
\label{coef11}
\beta_0 (k) \equiv \langle \hat{\psi} (\cdot, k ), \varphi_0 (\cdot; k)
\rangle,
\eeq
with $\hat{\psi}$ denoting the partial Fourier transform given by
\beq
\label{partft1}
\hat{\psi} (x,k ) \equiv \frac{1}{\sqrt{2 \pi}} \int_\R ~ e^{-iky} \psi(x,y)
~dy.
\eeq
The matrix element of the current operator $V_y$ in such a state is
\beq
\label{curr1}
\langle \psi , V_y  \psi \rangle = \int_\R ~dx \int_{\omega_0^{-1}
(\Delta_0)} ~dk | \beta_0 (k)|^2 (k - Bx) |
\varphi_0 (x;k)|^2 .
\eeq
From (\ref{ev1}) and the Feynman-Hellmann Theorem, we find that
\beq
\label{disp1}
\omega_0 ' (k) = 2 \int_{\omega_0^{-1} (\Delta_0)} ~dx
~(k - Bx) ~| \varphi_0 (x;k)|^2 ,
\eeq
so that we get
\beq
\label{curr2}
\langle \psi , V_y  \psi \rangle = \frac{1}{2}
\int_{\omega_0^{-1} ( \Delta )}  | \beta_0 (k)|^2 ~\omega_0 ' (k) ~dk.
\eeq
It follows from (\ref{curr2}) that in order to obtain a lower bound on the
expectation of the current operator in the state $\psi$ we need to bound the
derivative $\omega_0 ' (k)$ from below for $k \in \omega_0^{-1} ( \Delta_0
)$.  The next step of the proof involves relating the derivative
$\omega_0 ' (k)$ to the trace of the eigenfunction $\varphi_0 (x;k)$
on the boundary $x=0$.
For this, we use the formal commutator expression
\beq
\label{commutator1}
\hat{V}_y (k) \equiv
(k-Bx) = \frac{-i}{2B} [ p_x, h_0(k)] + \frac{1}{2B} V_0' (x).
\eeq
Inserting this into the identity (\ref{disp1}), we find
\bea
\label{derivomega}
\omega_0 ' (k) &=& 2 \langle \varphi_0 (\cdot; k ), (k-Bx) \varphi_0
(\cdot; k) \rangle \nonumber \\
&=& \frac{-i}{2B} \langle \varphi_0 (\cdot; k ),[ p_x, h_0(k)]
\varphi_0(\cdot; k) \rangle + \frac{-\mathcal{V}_0 }{B} \varphi_0 (0;k)^2
\nonumber \\
&=& \frac{-\mathcal{V}_0 }{B} \varphi_0 (0;k)^2,
\eea
since the commutator term vanishes by the Virial Theorem. Consequently, we
are left with the task of estimating the trace of the eigenfunction along
the boundary. Much of our technical work is devoted to obtaining lower
bounds on quantities of the form $\mathcal{V}_0 \varphi_n (0;k)^2$, for
$n=0,1, 2, \ldots $.
The situation for the two-edge geometries is more complicated since there
is an edge current associated with each edge.
This analysis of two-edge geometries is the subject of \cite{[HS2]}.

Let $H = H_L (B) + V_0 + V_1$ be a perturbation of
the one-edge Hamiltonian with spectral family $E ( \cdot )$.
We consider an energy interval $\Delta_n \subset
(E_n(B) , E_{n+1} (B) )$, and $| \Delta_n |$ small.
Roughly speaking, the main result of this paper is a uniform lower bound on
the
expectation of edge currents in all states with energy localized in the
interval
$\Delta_n$. We prove that for each $n \in \N$, there exists a finite
constant $C_n > 0$ (given precisely below), so that if
$\psi \in E ( \Delta_n ) L^2 ( \R^2)$,
and the perturbation $V_1$ is such that
$\| V_1 \|_\infty / B$ is sufficiently small, then
\beq
\label{mainresult}
| \langle \psi , V_y \psi \rangle | \geq C_n B^{1/2} \| \psi \|^2 .
\eeq

We note that the order $B^{1/2}$ in (\ref{mainresult}) is optimal as for
the unperturbed model, we prove that
\beq
\label{mainresult2}
C_n B^{1/2} \| \psi \|^2 \leq | \langle \psi, V_y \psi \rangle | \leq (1/
C_n) B^{1/2} \| \psi \|^2.
\eeq
We make two remarks about this result, one concerning the time-dependent
theory,
and the second concerning the IQHE.
First, we remark that
the time-independent estimate (\ref{mainresult}) implies that
the current persists with at least the same
strength for all times provided that the bulk Hamiltonian $H_{bulk} = H_L
(B) + V_1$ has a gap in its spectrum between the Landau levels.
That is, the estimate
(\ref{mainresult}) remains the same if we replace
$\psi$ with $\psi_t = e^{-iH t} \psi$,
or, equivalently, if we replace the current
operator $V_y$ with the Heisenberg current operator $V_y (t)
= e^{-iHt} V_y e^{iHt}$.
The edge current also remains localized in a neighborhood of
size $\mathcal{O} (B^{-1/2})$ near the boundary for all time.
Secondly, it has recently been proved that the conductivity
corresponding to the edge current,
called the {\it edge conductivity} $\sigma_e$, is quantized, and, in fact,
equal to the bulk conductivity, $\sigma_b$.
The edge currents studied in this paper correspond to the edge conductivity
and we refer to the papers
\cite{[CG],[CGH],[EG],[EGS],[KS1],[KS2],[KRS],[SKR]}.
For the importance of edge currents in the IQHE, we refer to the
papers \cite{[Hal], [HT], [KRS]}.

\subsection{Related Papers}

There are several papers on the
subject of edge currents for unbounded, one-edge geometries.
Macris, Martin, and Pul\'e \cite{[MMP]} studied the half-plane
case of one straight edge
with {\it soft} confining potentials. We extend this work
proving the existence
of edge currents for a large family of soft confining potentials
in section 6. Furthermore, we
show that we can interpolate between soft and hard confining potentials.
DeBi\`evre and Pul\'e \cite{[DBP]}
considered the case of a {\it hard} confining potential, that is,
Dirichlet boundary conditions (DBC). We treat this case
in sections 3 and 5 and show that
show that one
can interpolate between soft and hard confining potentials.
The case of DBC was also treated by
Fr\"ohlich, Graf, and Walcher \cite{[FGW]} who studied
non-straight edges. We consider non-straight edges in section 4.
As explained in section
5, these papers \cite{[DBP],[FGW],[MMP]}
linked the spectral properties of the one-edge Hamiltonians to
the existence of edge currents through the use of the Mourre commutator
method.
We discuss this thoroughly in section 5.
The main interest in spectral properties
is due to the fact that these authors prove that under weak perturbations
(relative to $B$) there is absolutely continuous spectrum in the
intervals $\Delta_n$.
It was pointed out by Exner, Joye, and Kova\v{r}\'ik \cite{[EJK2]}
that absolutely continuous spectrum and edge currents can appear when the
edge is simply an infinite array of point interactions. These authors
studied the Hamiltonian (1.2) for which $V_0(x) = \sum_{j \in \Z} \alpha
\delta (x - j )$, and proved that there are bands of absolutely continuous
spectra between the Landau levels and that the Landau levels remain
infinitely degenerate.
More recently, Buchendorfer and Graf \cite{[BG]}
developed a scattering theory for edge
states in one-edge geometries. These authors show that edge states acquire
a phase due to a bend in the boundary relative to a state propagating along
a straight boundary. This work has some similarities with the material
in section 4.

\subsection{Contents}

The content of this paper is as follows. Section 2 is devoted the proofs of
the edge current estimates for the case
of a Sharp Confining Potential and a straight edge.
In section 3, we extend these results to the case of Dirichlet boundary
conditions along the straight edge.
Section 4 is devoted to considering more general boundaries. We introduce
the notion of asymptotic edge currents and use scattering theory
to prove the stability of these currents. Spectral properties of the
Hamiltonians associated with one-edge geometries are studied in section 5
using the Mourre commutator method.
In section 6, we extend the results to soft confining potentials.
The paper concludes with three appendices. The first appendix in section 7
presents results on the dispersion curves needed in the proofs. The second
appendix in section 8, of independent interest, provides explicit pointwise
upper and lower bounds on solutions to a certain form
of second-order ODEs.
In appendix 3, section 9,
we apply these results to obtain eigenfunction bounds for
our specific operators.

\subsection{Acknowledgments}

We thank J.-M.\ Combes for many discussions on edge currents and their
role in the IQHE. We also thank F.\ Germinet, G.-M.\ Graf,
E.\ Mourre, and H.\
Schulz-Baldes for fruitful
discussions. Some of this work was done when
ES was visiting the Mathematics Department
at the University of Kentucky and he thanks the Department for its
hospitality and support.


\renewcommand{\thechapter}{\arabic{chapter}}
\renewcommand{\thesection}{\thechapter}

\setcounter{chapter}{2} \setcounter{equation}{0}

\section{The Straight Edge and a Sharp Confining Potential}

In this section, we prove an explicit lower-bound on the edge current formed
by
a sharp confining potential $V_0 (x) = \mathcal{V}_0 \chi_{\{x<0\}} (x)$
along the straight edge $x = 0$.
The nonperturbed, one-edge geometry Hamiltonian $H_0
= H_L(B) + V_0$, is a nonnegative, self-adjoint operator on $D(H_L(B))$.
We write $E_0 ( \cdot )$ for the spectral family of $H_0$.
If a classical electron has energy below $\mathcal{V}_0$, then the
corresponding classical Hamiltonian describes the dynamics of the particle
in
the half-plane $x > 0$, the classically allowed region. The
complementary region is the classically forbidden region for an electron
with energy less than $\mathcal{V}_0$.
The edge $x=0$ reflects the cyclotron orbits of these electrons
and causes a net drift of the electron along the edge.
This is the origin of the edge current.
We will later treat a general
family of perturbations $V_1$, and prove the
persistence of edge currents, provided $\| V_1 \|_\infty$ is small enough
relative to $B$ (and without assuming that $V_1$ is differentiable as
required by some commutator methods).
As discussed in section 5,
similar results for more restrictive potentials $V_1$
can be derived from commutator estimates, as
obtained by DeBi\`evre and Pul\'e
\cite{[DBP]}, and by Fr\"ohlich, Graf, and Walcher \cite{[FGW]}.

\subsection{The Main Results for the Unperturbed Case}

Our main result is an explicit lower-bound on the size of the edge current
for half-plane in certain states for the unperturbed Hamiltonian $H_0$.
In order to formulate the main theorem, we need to describe these states.
Because the edge is straight, we can use the Fourier transform with respect
to the $y$-variable to reduce the problem to a one-dimensional one.
The unperturbed operator $H_0$ admits a partial Fourier decomposition with
respect to the $y$-variable, and the Hilbert space $L^2 ( \R^2 )$ can be
expressed as a constant fiber direct integral over $\R$ with fibers $L^2 (
\R)$. For $H_0$, we write
\beq
\label{dirsum1}
H_0 = \int^{\oplus}_{\R} h_0(k) ~dk ,
\eeq
where
\beq
\label{fibered1}
h_0(k) = p_x^2 + (k-Bx)^2 + V_0(x), ~~\mbox{on} ~~L^2 ( \R).
\eeq
As in section 1, we write $\varphi_j (x;k)$ and $\omega_j (k)$ for the
normalized eigenfunctions and the corresponding eigenvalues.
The eigenvalues are nondegenerate (cf.\ section 7) and, consequently, we
choose the eigenfunctions $\varphi_j$ to be real. These eigenfunctions
form an orthonormal basis of $L^2 ( \R )$, for any $k \in \R$.
Because the map $k \rightarrow h_0(k)$ is operator analytic,
the simple eigenvalues $\omega_j(k)$ are analytic functions of $k$.
We are interested in states that are energy localized in intervals
$\Delta_n$ lying between two consecutive Landau levels, that is
$\Delta_n \subset ( E_n (B) , E_{n+1} (B) )$.
Consider a state $\psi$ having the property that $\psi = E_0 (
\Delta_n ) \psi$. For such a state $\psi$, we can take the
Fourier transform of $\psi$
with respect to $y$ and, using an eigenfunction expansion,
write
\beq
\label{decomp1}
\psi ( x, y) = \frac{1}{\sqrt{2 \pi}} \Sum_{j=0}^n
\int_\R  e^{iky} ~\chi_{ \omega_j^{-1}(
\Delta_n )} (k) \beta_j (k) \varphi_j (x;k) ~dk ,
\eeq
where the coefficients $\beta_j (k)$ are defined by
\beq
\label{beta1}
\beta_j (k) \equiv \langle \hat{\psi} (\cdot, k ),
\varphi_j ( \cdot ; k) \rangle,
\eeq
where the partial Fourier transform is defined in
(\ref{partft1}).
The normalization is such
\beq
\label{stnorm1}
\| \psi\|_{L^2(\R^2)}^2 = \Sum_{j=0}^n  \int_{\omega_j^{-1} (\Delta_n)} |
\beta_j (k) |^2 ~dk.
\eeq
Throughout the paper, we will take
the interval $\Delta_n \subset ( E_n(B), E_{n+1} (B))$ to be given by
\beq
\label{interval1}
\Delta_n = [ (2n+a)B, (2n+c)B], ~~\mbox{for} ~~ 1 < a < c < 3.
\eeq
We can now state the main theorem for the unperturbed, single straight edge
Hamiltonian $H_0$ with a sharp confining potential.

\vspace{.1in}
\noindent
\begin{theorem}
For $n \geq 0$,
let $\Delta_n$ be as in (\ref{interval1}),
and suppose that
$\mathcal{V}_0 > (2n+3)B$.
Let $E_0 (\Delta_n)$ be the spectral projection for $H_0$ and
the interval $\Delta_n$. Let
$\psi \in L^2 (\R^2)$ be a state satisfying $\psi = E_0 (
\Delta_n ) \psi$ with an expansion as in (\ref{decomp1})--(\ref{stnorm1}).
Then, for $c-a > 0$ sufficiently small, if $n \geq 1$,
so that condition (\ref{empty1}) is
satisfied, we have,
\bea
\label{current3}
- \langle \psi, V_y \psi \rangle &\geq &
\frac{1}{2^4
(n+1)^2 [\mathcal{H}^{(n)}]^2} \left( \frac{\pi}{B^7} \right)^{1/2}
\Sum_{j=0}^n \int_{\omega_j^{-1} ( \Delta_n )} | \beta_j(k)|^2
   \nonumber \\
& & \times  \left( 1 - \frac{\omega_j(k)}{\mathcal{V}_0} \right)
( \omega_j (k) - E_n (B) )^2 ( E_{n+1} (B)) - \omega_j (k) )^2
~dk,\nonumber \\
& &
\eea
where the constant $\mathcal{H}^{(n)}$ is defined in (\ref{defn1}).
\end{theorem}

\vspace{.1in}
\noindent
Let us note a simplification of the above expression under
reasonable conditions.
For $k \in \omega_j^{-1} ( \Delta_n)$, $j=0, \ldots, n$, we have
\beq
\label{inclusions1}
E_n(B) < (2n + a)B \leq \omega_j (k) \leq (2n+3)B < E_{n+1} (B),
\eeq
we have
\beq
\label{inclusions2}
(\omega_j (k) - E_n (B) )^2 =  B^2 (a -1)^2, ~~( E_{n+1} (B) - \omega_j
(k) )^2 =  B^2 (3-c)^2.
\eeq

\vspace{.1in}
\noindent
\begin{corollary}
Let us suppose that
$\mathcal{V}_0 > (2n+3) B$, for $n \geq 0$,
is such that for $k \in \omega_j^{-1} (
\Delta_n)$, we have
\beq
\left( 1 - \frac{\omega_j(k)}{\mathcal{V}_0} \right) > \frac{1}{2}.
\eeq
Then, under this condition, the hypotheses of Theorem 2.1,
and recalling (\ref{inclusions2}), the edge current satisfies the bound
\beq
\label{current4}
- \langle \psi, V_y \psi \rangle \geq \frac{ \pi^{1/2} (a-1)^2
  (3-c)^2}{2^5 (n+1)^2
[\mathcal{H}^{(n)}]^2} ~B^{1/2} \| \psi \|^2.
\eeq
Note that for $n=0$, the constant $\mathcal{H}^{(0)} = 1$.
\end{corollary}

This result shows that any state with energy between $E_n(B)$ and
$E_{n+1}(B)$
carries an edge current. However, as the
energy approaches a Landau level, the state may delocalize away from the
edge.

\subsection{Proof of Theorem 2.1.}

In order to prove Theorem 2.1,
we note that
from the representation (\ref{decomp1}), the matrix element of the edge
current can be written as
\bea
\label{curr3}
\lefteqn{ \langle \psi , V_y \psi \rangle} \nonumber \\
& = &
\Sum_{j,l=0}^n \int_\R  ~\chi_{\omega_l^{-1} (\Delta_n)}(k)
~\chi_{ \omega_j^{-1}(\Delta_n )} (k) \overline{\beta}_l (k) \beta_j (k)
\langle  \varphi_l (\cdot;k), (k-Bx) \varphi_j ( \cdot; k ) \rangle
~dk \nonumber \\
&=& \mathcal{M}_n (\psi) + \mathcal{E}_n (\psi) ,
\eea
where the main term $\mathcal{M}_n ( \psi )$ is given by
\beq
\label{curr4m}
\mathcal{M}_n ( \psi ) \equiv
\Sum_{j=0}^n \int_\R  ~\chi_{ \omega_j^{-1}(\Delta_n )} (k)
   | \beta_j (k) |^2 \langle \varphi_j (\cdot;k), (k-Bx) \varphi_j
     ( \cdot; k ) \rangle ~dk .
\eeq
The term $\mathcal{E}_n (\psi)$ is the error term involving the
cross-terms between different Landau levels. It is given by
\beq
\label{error1}
\mathcal{E}_n (\psi) \equiv \Sum_{j \neq l; j,l =0}^n \int_{\R}  ~\chi_{\omega_l^{-1} (\Delta_n)}(k)
~\chi_{ \omega_j^{-1}(\Delta_n )} (k) \overline{\beta}_l (k) \beta_j (k)
\langle  \varphi_l (\cdot;k), (k-Bx) \varphi_j ( \cdot; k ) \rangle ~dk. 
\eeq
Concerning this term, we have the following result.

\vspace{.1in}
\noindent
\begin{lemma}
Suppose  $\Delta_n \subset (E_n (B) , E_{n+1} (B))$
has the form given in (\ref{interval1}).
Under the conditions described above, if $c-a$ is sufficiently small so that
condition (\ref{empty1}) is satisfied, then the error term
(\ref{error1}) for the unperturbed problem with any $0 \leq \mathcal{V}_0 <
\infty$ is zero: $\mathcal{E}_n (\psi) = 0$.
\end{lemma}

\vspace{.1in}
\noindent
{\bf Proof.}
The vanishing of $\mathcal{E}_n (\psi)$ follows from the fact that
$\sigma_{jl} \equiv \omega_l^{-1} (\Delta_n) \cap \omega_j^{-1}
(\Delta_n) = \emptyset$, for $j \neq l$
and for $|\Delta_n|$ sufficiently small.
Each dispersion curve $\omega_j (k)$ is strictly monotone decreasing as
follows from the representation (\ref{derivomega}),
together with the formula in
Proposition 2.1 and the bound in Lemma 2.3. Furthermore, the dispersion
curves never intersect. For suppose that there exists a
$k_0$ so that $\omega_j (k_0) = \omega_l (k_0)$, for some $j \neq l$. This
means that $h_0 (k_0)$ has a doubly-degenerate eigenvalue, a contradiction
to the simplicity of the spectrum of $h_0 (k)$ (cf.\ Proposition 7.2).
Let us suppose that $\omega_j (k) < \omega_l (k)$,
and let $k_l^c$ be the unique point satisfying
$\omega_l (k ) = (2n+c)B$.
Now, it is easy to check that the condition
that guarantees that $\sigma_{jl} = \emptyset$ is that
\beq
\label{empty1}
( (2n+c)B - \omega_j (k_l^c)) > (c-a)B.
\eeq
Since the right side of (\ref{empty1}) can be made small be taking $a$ close
to $c$, whereas
the left side is independent of $a$, this proves the result. $\Box$ \\

\noindent
We note that even when the sets $\sigma_{jl}$ are nonempty, the
eigenfunctions of the reduced Hamiltonians $h_0(k)$ are spatially localized
so that the error term $\mathcal{E}_n (\psi)$ is exponentially small.

We therefore have to
estimate the main term in (\ref{curr3}).
It is clear that we need to control the matrix
element of $\hat{V}_y = (k - Bx)$ in the states $\varphi_j (x; k)$.
The following formal commutator expression plays an important role in the
calculation of the current in these eigenstates:
\beq
\label{comm1}
\hat{V}_y = (k-Bx) \equiv \frac{-i}{2B} [ p_x , h_0 (k) ] +
\frac{1}{2B} V_0' ,
\eeq
where $V_0'$ is interpreted in the distributional sense.
As a first step, we note
the following basic result that follows from analyticity,
the Virial Theorem, the
existence of $\varphi_j (0 ; k)$ as proved in Proposition 7.1, and the
expression (\ref{comm1}).

\vspace{.1in}
\noindent
\begin{proposition}
Let $\varphi_j (x;k)$ be an eigenfunction of $h_0
(k)$, with eigenvalue $\omega_j(k)$.
We have
\beq
\label{current6}
\langle \varphi_j ( \cdot; k) , \hat{V}_y \varphi_j ( \cdot;k) \rangle =
- \frac{\mathcal{V}_0 }{2B} \varphi_j ( 0; k )^2 .
\eeq
\end{proposition}

\vspace{.1in}

Recall that the matrix element
in (\ref{current6}) is equal to $(1/2) \omega_j ' (k)$.
So the problem is to estimate the slope $\omega_j ' (k)$ of the dispersion
curves from below for $k \in \omega_j^{-1} ( \Delta_n )$, for $j = 1 ,
\ldots, n$.
In light of this estimate, the main term of the edge current in
(\ref{curr3}) can be written as
\beq
\label{main1}
\mathcal{M}_n ( \psi)
\equiv - \frac{1 }{2B} \Sum_{j=0}^n \int_{\omega_j^{-1}(\Delta_n )}
| \beta_j (k) |^2 \; ( \mathcal{V}_0 \varphi_j ( 0; k )^2  ) ~dk.
\eeq
Our next step is to obtain a lower bound on the trace of the eigenfunction
on
the edge, so as to be able to estimate $\mathcal{V}_0 \varphi_j ( 0; k )^2$
from below. This will require several steps.

\vspace{.1in}
\noindent
{\bf STEP 1: Eigenfunction Estimate}

For the normalized real eigenfunction $\varphi_j ( x ;k)$,
we define, for any $\delta \geq 0$,
\beq
\label{etaL}
\eta_j ( \delta ) \equiv \varphi_j ( - \delta ; k )^2 .
\eeq
We now obtain exponential decay results on $\eta_j ( \delta )$ as $\delta
\rightarrow \infty$.
An ODE method allows one to obtain a precise form of the prefactor.

\vspace{.1in}

\noindent
\begin{theorem}
Let $\varphi_j (x;k)$
be the normalized real eigenfunction of $h_0(k)$, defined above,
with corresponding eigenvalue $\omega_j (k)$.
Then, for any $\delta > 0$, and
for all $k \in \R$ so that $0 \leq \omega_j(k) < \mathcal{V}_0$, we have
\beq
\label{efdecay2}
\eta_j ( \delta) \leq \eta_j (0) e^{- \sqrt{2 (\mathcal{V}_0 -
\omega_j(k))} \delta } .
\eeq
\end{theorem}

\vspace{.1in}

\noindent
{\bf Proof.}

\noindent
1.  The idea of the proof is to obtain a good lower bound on ${\eta''}_j (
\delta)$ and to integrate the result. We refer the reader to appendix 1,
Proposition 7.1, on the differentiability of $\varphi_j (x;k)$.
The first derivative of $\eta_j ( \delta)$ with respect to $\delta$
is easily computed:
\bea
\label{eq:firstder}
\eta_j'(\delta) & = & -2 \partial_x \varphi(-\delta;k)
  \; \varphi(-\delta;k)  \nonumber \\
& = & -2 \left[ \int_{-\infty}^{-\delta} (\partial_t^2 \varphi ) (t;k)
      \; \varphi(t;k) dt \right. \nonumber \\
& & \left. + \int_{-\infty}^{-\delta} (\partial_t \varphi )(t;k)^2 dt
\right].
\eea
We use the eigenvalue equation $h_0(k) \varphi_j = \omega_j(k) \varphi_j$
to re-express $\partial_t^2 \varphi_j$ for $t < 0$ as
\beq
\label{ev2}
\partial_t^2 \varphi_j(t;k) =
   (k- Bt )^2 \varphi_j (t;k)  + (\mathcal{V}_0 - \omega_j(k) ) \varphi_j
(t;k).
\eeq
Substituting this into (\ref{eq:firstder}), we obtain,
\bea
\label{deriv1}
  -\demi \eta_j'(\delta) & = & (\mathcal{V}_0 - \omega_j(k) )
   \int_{-\infty}^{-\delta} \varphi_j (t;k)^2 dt \nonumber \\
   & & + \int_{-\infty}^{-\delta} ( \partial_t \varphi_j )(t;k)^2 dt +
    \int_{-\infty}^{-\delta} (k- Bt )^2 \varphi_j (t;k)^2 ~dt.
\eea

\noindent
2. We now take the derivative with respect to $\delta$ of the
terms in (\ref{deriv1}). This gives
\bea
\label{eq:secondder}
\demi \eta_j''(\delta) & = & (\mathcal{V}_0 - \omega_j(k))
\eta(\delta)  \nonumber \\
& &  + ( \partial_x \varphi_j )(-\delta;k)^2 +
  (k+B\delta )^2 \varphi_j (-\delta;k)^2.
\eea
Since the last two terms
on the right of (\ref{eq:secondder}) are nonnegative, we
have proved the lower bound
\begin{equation}
\label{InEDO}
\eta_j''(\delta) \geq 2(\mathcal{V}_0 - \omega_j(k) )
\eta_j(\delta).
\end{equation}
As $\eta_j'$ obviously converges to zero at infinity, it follows
from
(\ref{InEDO}) that $\eta_j'(\delta) \leq 0$ for any $\delta \in
\R_+$.
So multiplying (\ref{InEDO}) by $\eta_j'(\delta)$ and integrating
along $[t,+\infty)$ for any $t \geq 0$
also gives :
\[  {\eta_j'}^2(t) \geq 2(\mathcal{V}_0  - \omega_j(k) )
\eta_j^2(t).
\]
By integrating along $[0,\delta]$, for any $\delta \geq 0$, one finally
obtains
\begin{equation}
\label{expodecay}
\eta_j(\delta) \leq \eta_j(0) \expo{-\sqrt{2(\mathcal{V}_0
- \omega_j(k))}
\delta},
\end{equation}
proving the result. $\Box$

\vspace{.1in}
\noindent
{\bf STEP 2: Harmonic Oscillator Eigenfunction Comparison}

It is useful to compare the eigenfunctions of $h_0(k)$ to those of the
harmonic oscillator Hamiltonian with no confining potential.
The harmonic oscillator Hamiltonian $h_B(k)$ on $L^2 ( \R)$ is
defined as
\beq
\label{harmosc1}
h_B(k) \equiv p_x^2 + (k - Bx)^2 .
\eeq
The eigenvalues of this operator are precisely the Landau energies $E_m(B)$
and are nondegenerate and independent of $k$. We will denote the real
normalized
eigenfunctions by $\psi_m (x;k)$. These are given by
\beq
\label{harmoscef1}
\psi_m (x;k) = \frac{1}{\sqrt{2^m m!}} \; \left( \frac{B}{\pi}
\right)^{1/4} \; e^{- \frac{B}{2} ( x- k/B)^2} H_m (x \sqrt{B} - (k /
\sqrt{B}) ),
\eeq
where $H_m(u)$ is the normalized Hermite polynomial with $H_0 (u) = 1$.
We expand the eigenfunctions $\varphi_j (x;k)$ in terms of these
eigenfunctions
\beq
\label{expand2}
\varphi_j (x;k) = \Sum_{m=0}^\infty \alpha_m^{(j)} (k) \psi_m (x;k),
\eeq
where the coefficients are given by
\beq
\label{coef1}
\alpha_m^{(j)}(k) = \langle \varphi_j (\cdot;k), \psi_m (\cdot; k) \rangle,
\eeq
and satisfy
\beq
\label{norm1}
\| \varphi_j (\cdot;k)\|^2 = \Sum_{m=0}^\infty |\alpha_m^{(j)}(k)|^2 = 1.
\eeq
We occasionally suppress the variable $k$ in the notation
and write $\alpha_m^{(j)}$ for these coefficients.

\vspace{.1in}

\noindent
\begin{lemma}
Let $P_n (k)$ be the projection on the eigenspace spanned by the
first $n$ eigenfunctions $\psi_m$ of the harmonic oscillator Hamiltonian
$h_B (k)$
(\ref{harmosc1}).
Let $\alpha_m^{(j)}$ be the expansion coefficients defined in (\ref{coef1}).
For all $k \in \omega_j^{-1} (\Delta_n)$, with $\Delta_n$ as in
(\ref{interval1}), and for all $0 \leq j \leq n$, we have
\beq
\label{lower2}
\Sum_{m=0}^n |\alpha_m^{(j)}(k)|^2 \geq \frac{1}{2B(n+1)} ( E_{n+1}(B) -
\omega_j (k) ) > 0 ,
\eeq
and
\beq
\label{lowerbound}
| \langle \varphi_j ( \cdot;k), V_0 P_n (k) \varphi_j (\cdot;k) \rangle|
\geq
\frac{1}{2B(n+1)} ( \omega_j(k) - E_n(B) )( E_{n+1}(B) -
\omega_j(k)) > 0.
\eeq
\end{lemma}

\vspace{.1in}

\noindent
{\bf Proof.}

\noindent
1. We compute the matrix element $\langle \varphi_j, V_0 \varphi_j \rangle$
using the expansion (\ref{expand2}),
\bea
\label{expand3}
\langle \varphi_j, V_0 \varphi_j \rangle &=&
\langle \varphi_j, ( h_0(k) -h_B(k) ) \varphi_j \rangle \nonumber \\
&=& \Sum_{m \geq 0} ( \omega_j (k) - E_m (B))  |\alpha_m^{(j)}(k)|^2 ,
\eea
using the normalization (\ref{norm1}). Rearranging the terms in
(\ref{expand3}), we find
\bea
\label{expand4}
\Sum_{m \leq n} (\omega_j (k) - E_m (B))  |\alpha_m^{(j)}(k)|^2 &=& \langle
\varphi_j, V_0 \varphi_j \rangle \nonumber \\
&& + \Sum_{m \geq n+1} (E_m(B) - \omega_j (k) )  |\alpha_m^{(j)}(k)|^2
\nonumber \\
& \geq & ( E_{n+1} (B) - \omega_j (k) ) \left( 1- \Sum_{m \leq n}
|\alpha_m^{(j)}(k)|^2 \right) . \nonumber \\
&&
\eea
We now assume that $k \in \omega_j^{-1} (\Delta_n)$ and $j \leq n$.
In this case, the coefficient $E_{n+1} (B) - \omega_j (k) > 0$. Moving the
second term on the right of (\ref{expand4}) to the left, we obtain
\bea
\label{expand5}
(E_{n+1} (B) - \omega_j (k) ) & \leq &
\Sum_{m \leq n} (\omega_j (k) - E_m (B) + E_{n+1} (B) - \omega_j (k))
|\alpha_m^{(j)}(k)|^2 \nonumber \\
&=& \Sum_{m \leq n} ( E_{n+1} (B)- E_m (B)) |\alpha_m^{(j)}(k)|^2 \nonumber
\\
&\leq&
2(n+1)B \left( \Sum_{m \leq n} |\alpha_m^{(j)}(k)|^2 \right).
\eea
The result (\ref{lower2}) follows from (\ref{expand5}).

\noindent
2. The calculation of $\langle \varphi_j ( \cdot;k), V_0 P_n (k) \varphi_j
(\cdot;k) \rangle$, for $k \in \omega_j^{-1} (\Delta_n)$, is similar. We
write
\bea
\langle \varphi_j ( \cdot;k), V_0 P_n(k) \varphi_j (\cdot;k) \rangle &=&
\langle \varphi_j ( \cdot;k), (h_0(k)-h_B(k)) P_n (k) \varphi_j (\cdot;k)
\rangle \nonumber \\
&=& \Sum_{m \leq n} (\omega_j (k) - E_m (B))  |\alpha_m^{(j)}(k)|^2
\nonumber
\\
&\geq& (\omega_j (k) - E_n (B)) \Sum_{m \leq n}|\alpha_m^{(j)}(k)|^2
\nonumber \\
& \geq & \frac{1}{2B(n+1)}
  (\omega_j (k) - E_n (B))( E_{n+1} (B) - \omega_j (k) ) ,
  \nonumber \\
\eea
where we used (\ref{lower2}). $\Box$

\vspace{.1in}

\noindent
{\bf STEP 3: Lower Bound on the Trace}

We now use the eigenfunction estimate of Step 1 and the lower bound of Step
2 in order to express the matrix element
$\langle \varphi_j ( \cdot;k), V_0 P_n(k) \varphi_j (\cdot;k) \rangle$ in
terms of the trace of $\varphi_j$ on the edge.
We recall that $P_n (k)$
is the projection onto the eigenspace spanned by the first $n$
eigenfunctions
of the harmonic oscillator Hamiltonian $h_B (k)$.
\vspace{.1in}

\noindent
\begin{lemma}
Let $\varphi_j (x;k)$ be an eigenfunction of $h_0(k)$, as above, for $0 \leq
j \leq n$.
Then, for all $k \in \omega_j^{-1} (\Delta_n)$, we have
\bea
\label{trace2}
\lefteqn{ \mathcal{V}_0^2 \varphi_j (0;k)^2 } \nonumber \\
&\geq& \left( \frac{\pi}{B} \right)^{1/2}
\frac{  [\mathcal{V}_0 - \omega_j(k)] }{8 B^2 (n+1)^2
  [ \mathcal{H}^{(n)} ]^2 }
(\omega_j (k) - E_n (B))^2 ( E_{n+1} (B) - \omega_j (k) )^2 ,  \nonumber \\
& &
\eea
where $\mathcal{H}^{(n)}$ is defined in (\ref{defn1}).
\end{lemma}

\noindent
{\bf Proof.}
We use the expansion of $\varphi_j$ in the eigenfunctions $\psi_m$ and
obtain
\bea
\label{expand6}
\langle \varphi_j ( \cdot;k), V_0 P_n (k) \varphi_j (\cdot;k) \rangle &=&
\mathcal{V}_0 \int_{-\infty}^0 \varphi_j ( x;k) P_n (k)
\varphi_j ( x;k) ~dx \nonumber \\
&=& \Sum_{m \leq n} \mathcal{V}_0 \alpha_m^{(j)}(k) \int_{-\infty}^0
\varphi_j ( x;k) \psi_m(x;k) ~dx . \nonumber \\
&&
\eea
To estimate the integral, we use
the exponential decay of the eigenfunctions $\varphi_j$ as given in
Theorem 2.2. For $x < 0$, the main eigenfunction decay estimate
(\ref{efdecay2}) gives
\beq
\label{expupper1}
\varphi_j (x;k)^2 \leq \varphi_j (0;k)^2 e^{- \sqrt{2 (\mathcal{V}_0 -
\omega_j(k))} |x| } .
\eeq
We recall that $\psi_m(x;k)$ is given in (\ref{harmoscef1}), and
define coefficients $C_m(B)$ and $\mathcal{H}_m$ by
\beq
\label{defn0}
C_m(B) \equiv \left( \frac{B }{ \pi} \right)^{1/4} ( 2^m m!)^{-1/2},
\mbox{and} ~\mathcal{H}_m  \equiv \sup H_m(u) e^{- u^2 / 2}.
\eeq
In terms of these coefficients, the integral can be bounded above by
\bea
\label{integral1}
\left| \int_{-\infty}^0 \varphi_j ( \cdot;k) \psi_m(x;k) ~dx \right|
& \leq & C_m (B)|\varphi_j (0;k )| \mathcal{H}_m
   ~\int_{0}^\infty  e^{- \sqrt{2^{-1} (\mathcal{V}_0 -
\omega_j(k))}x} ~dx  \nonumber \\
&\leq& \frac{ 2^{1/2} C_m (B) |\varphi_j (0;k )| \mathcal{H}_m }{
    \sqrt{ ( \mathcal{V}_0 -\omega_j(k)) }} .
\eea
From (\ref{expand6}) and (\ref{integral1}), we get
\beq
\label{expand7}
| \langle \varphi_j ( \cdot;k), V_0 P_n (k) \varphi_j (\cdot;k) \rangle |
  \leq  \left( \frac{B}{\pi} \right)^{1/4}
    \frac{ 2^{1/2} \mathcal{V}_0 | \varphi_j (0;k) |}{ \sqrt{ (
\mathcal{V}_0
-\omega_j(k))} } \left( \Sum_{m \leq n} \frac{1}{\sqrt{ 2^m m!}}
\mathcal{H}_m | \alpha_m^{(j)}(k) | \right) .
\eeq
We define a constant $\mathcal{H}^{(n)}$ by
\beq
\label{defn1}
\mathcal{H}^{(n)} \equiv \left( \Sum_{m \leq n} \frac{1}{ 2^m m!}
\mathcal{H}_m^2 \right)^{1/2}, ~\mbox{where} ~\mathcal{H}_m \equiv \sup_{u
\in \R} H_m(u) e^{- u^2 / 2}.
\eeq
Applying the Cauchy-Schwarz inequality to the sum in (\ref{expand7}), and
recalling the normalization (\ref{norm1}), we find that
\beq
\label{expand8}
| \langle \varphi_j ( \cdot;k), V_0 P_n (k) \varphi_j (\cdot;k) \rangle |
  \leq  \left( \frac{B}{\pi} \right)^{1/4} \frac{ 2^{1/2} \mathcal{V}_0 |
\varphi_j (0;k) | \mathcal{H}^{(n)}}{ [ \mathcal{V}_0
-\omega_j(k) ]^{1/2}  }.
\eeq
We square expression (\ref{expand8}), and use the bound (\ref{lowerbound})
in Lemma 2.4, to obtain the result (\ref{trace2}). $\Box$

\vspace{.1in}

The proof of Theorem 2.1
now follows directly from the expression for the main term $\mathcal{M}_n
(\psi)$
in (\ref{main1}) and the
lower bound for the expression $\mathcal{V}_0 \varphi_j (0;k)^2$ given
in Lemma 2.3. Corollary 2.1
follows directly from the lower bound on the main term.

\subsection{Perturbation Theory for the Straight Edge}

We now consider the perturbation of $H_0$ by a bounded potential
$V_1(x,y)$.
We prove that the lower bound on the edge current is stable with respect to
these perturbations provided $\| V_1 \|_\infty$ is not too large
compared with $B$.
As above, let $\Delta_n \subset ( E_n (B) , E_{n+1} (B))$ be a closed,
bounded
interval as in (\ref{interval1}) determined by the constants $1 < a < c <
3$.
We consider a larger interval $\tilde{\Delta}_n$, containing $\Delta_n$,
with the same midpoint $E_n = (2n + (a+c) \slash 2 )B \in  \Delta_n$,
and of the form
\beq
\label{outint1}
\tilde{\Delta}_n = [ (2n + \tilde{a})B, (2n+\tilde{c})B], ~~\mbox{for}
~1 < \tilde{a} < a < c < \tilde{c} < 3.
\eeq
In this perturbation argument, we calculate the velocity $V_y$ in states
$\psi \in E ( \Delta_n ) L^2 ( \R^2)$ that are close to states in $E_0
(\tilde{\Delta}_n ) L^2 ( \R^2)$. This closeness is measured by the constant
$\kappa > 0$ that we now define. First, we choose the constants
$\tilde{a}$ and $\tilde{c}$ in (\ref{outint1}) so that $\tilde{c} -
\tilde{a}$ is small enough for Theorem 2.1 to hold for states in
$E_0 (\tilde{\Delta}_n ) L^2 ( \R^2)$. Next, we choose a
constant $B_n > 0$ large enough and the constants $a$ and $c$, with
$c-a$ small enough, so that for all $B > B_n$, the constant $\kappa$
defined by
\beq
\label{defnkappa1}
\kappa^2 \equiv \left( 1 - \left( \frac{2}{\tilde{c} - \tilde{a} }
\right)^2 \left( \frac{ c-a}{2} + \frac{ \| V_1 \|_\infty}{B} \right)^2
\right) ,
\eeq
satisfies $0 < \kappa \leq 1$.

\vspace{.1in}
\noindent
\begin{theorem}
Let $V_1 (x,y)$ be a bounded potential
and let $E (\Delta_n)$ be the spectral projection for $H = H_0 + V_1$ and
the interval $\Delta_n$ as in (\ref{interval1}). Let
$\psi \in L^2 (\R^2)$ be a state satisfying $\psi = E (
\Delta_n ) \psi$. Let $\phi \equiv E_0 (
\tilde{\Delta}_n ) \psi$ and $\xi \equiv E_0 ( \tilde{\Delta}_n^c ) \psi$,
so that $\psi = \phi + \xi$.
Under the conditions given above on $a, c, \tilde{a}, \tilde{c}$, and for
$B > B_n$, the constant $\kappa$, defined
in (\ref{defnkappa1}), satisfies $0 < \kappa \leq 1$ and we have
\beq
\label{poskappa1}
\| \phi \| \geq \kappa \| \psi \| .
\eeq
Furthermore, we have the lower bound
\beq
\label{current1p}
- \langle \psi, V_y \psi \rangle \geq
B^{1/2} \kappa^2
~( C_n ( 3 - \tilde{c})^2 (\tilde{a} - 1)^2 - F( n, \| V_1 \| / B ) )
  ~\| \psi \|^2,
\eeq
where the constants are defined by
\beq
\label{ncnst1}
C_n = \frac{ \pi^{1/2}}{2^5 (n+1)^2 [ \mathcal{H}^{(n)}]^2} ,
\eeq
and
\bea
\label{ncnst2}
F(n, \| V_1 \|_\infty / B ) & = & (1-\kappa^2)^{1/4}
\left( 2n + c + \frac{ \|V_1 \|_\infty }{B} \right)^{1/2} \left( 2 +
  \sqrt{1 - \kappa^2} \right) \nonumber \\
& & + C_n ( 1 - \kappa^2 ) ( 3 - \tilde{c} )^2 (\tilde{a} - 1)^2  .
\eea
If we suppose that $\| V_1 \|_\infty < \mu_0 B$, then for a fixed level $n$,
if $c-a$ and $\mu_0$ are
sufficiently small (depending on $\tilde{a}$, $\tilde{c}$, and $n$),
there is a constant $D_n > 0$ so that
for all $B$, we have
\beq
\label{pertcurr1}
- \langle \psi, V_y \psi \rangle \geq
D_n \kappa^2 B^{1/2} \| \psi \|^2 .
\eeq
\end{theorem}

\noindent
{\bf Proof.}
With reference to the definitions (\ref{interval1}) and (\ref{outint1}),
we write the function $\psi$ as
\beq
\label{decomp1p}
\psi = E_0 (\tilde{\Delta}_n) \psi + E_0 (\tilde{\Delta}_n^c) \psi
\equiv \phi + \xi .
\eeq
We then have
\beq
\label{ident1}
\langle \psi, V_y \psi \rangle = \langle \phi, V_y \phi \rangle + 2 Re
\langle \phi, V_y \xi \rangle + \langle \xi, V_y \xi \rangle .
\eeq
The result follows from Theorem 2.1 provided we have a good bound on $\| \xi
\|$
and on $\| V_y \xi \|$. Let $E_n = (2n + (a+c) \slash 2)B  \in \Delta_n$ be
the midpoint of the intervals $\Delta_n$
and $\tilde{\Delta}_n$. We first note that
\bea
\label{xibound1}
\| \xi \| & \leq & \| E_0 (\tilde{\Delta}_n^c)(H_0 -E_n)^{-1} (H -E_n)\psi
\|
+ \| E_0 (\tilde{\Delta}_n^c)(H_0 -E_n)^{-1} V_1 \psi \|
\nonumber \\
&\leq & \frac{1}{d(E_n, \tilde{\Delta}_n^c)} \left( \frac{| \Delta_n|}{2} +
\| V_1 \| \right) \|
\psi \| \nonumber \\
& \leq & \left( \frac{2}{\tilde{c} - \tilde{a}} \right)
~\left( \frac{(c-a)}{2} + \frac{\|V_1
\|}{B} \right) \| \psi \| .
\eea
The bound (\ref{poskappa1})
follows from (\ref{xibound1}) and the orthogonality of $\phi$ and $\xi$.
Similarly, we find that
\bea
\label{vxibound1}
\| V_y \xi \|^2 &\leq& \langle \xi , H_0 \xi \rangle \nonumber \\
& \leq & | \langle \psi , H \xi \rangle | + \| V_1 \| ~\| \xi \| ~\| \psi \|
\nonumber \\
& \leq & ( (2n + c)B + \| V_1 \| ) ~\| \xi \| ~\| \psi \| .
\eea
Combining (\ref{xibound1}) and (\ref{vxibound1}), we obtain
\beq
\label{vxibound2}
| \langle \xi, V_y \xi \rangle | \leq \left( \frac{2}{\tilde{c} -
\tilde{a}} \right)^{3/2} ~B^{1/2}
~\left( \frac{(c-a)}{2} + \frac{ \| V_1 \|}{B} \right)^{3/2}  \left(
2n + c + \frac{\| V_1 \|}{B} \right)^{1/2} ~\| \psi \|^2 ,
\eeq
and
\beq
\label{vxibound3}
| \langle \phi, V_y \xi \rangle | \leq \left( \frac{2}{\tilde{c}
-\tilde{a}} \right)^{1/2} ~B^{1/2} ~\left( \frac{(c-a)}{2} + \frac{ \| V_1
\|}{B}
\right)^{1/2}  \left( 2n + c + \frac{\| V_1 \|}{B} \right)^{1/2}  ~\| \psi
\|^2.
\eeq
The lower bound on the main term in (\ref{ident1}) follows from
(\ref{current4}) of Corollary 2.1, and (\ref{outint1}),
\bea
\label{main2}
- \langle \phi , V_y \phi \rangle &\geq &\left( \frac{ \pi^{1/2} ( \tilde{a}
-
  1)^2 (\tilde{c} - 3)^2 }{2^5 (n+1)^2 [ \mathcal{H}^{(n)} ]^2 } \right)
~B^{1/2} ~\left( \Sum_{j=0}^n \int_{\omega_j^{-1} ( \tilde{\Delta}_n )}
| \beta_j (k)|^2 ~dk \right) \nonumber \\
&=& \left( \frac{ \pi^{1/2} ( \tilde{a} -1)^2 (\tilde{c} - 3)^2 }{2^5
(n+1)^2 [ \mathcal{H}^{(n)} ]^2 } \right)~B^{1/2} ( \| \psi \|^2 - \| \xi
\|^2 ) .
\eea
Combining this lower bound (\ref{main2}), with the estimate on $\| \xi \|$
in
(\ref{xibound1}), and the bounds (\ref{vxibound1})--(\ref{vxibound3}),
we find (\ref{current1p}) with the constants (\ref{ncnst1}) and
(\ref{ncnst2}). This completes the proof. $\Box$

We remark that if the state $\psi \in E ( \Delta_n) L^2 (\R^2)$ has the
property that the corresponding $\phi =0$, then the right side of
(\ref{main2}) is zero.
It follows from (\ref{poskappa1}), however,
that if the interval $\Delta_n$
is small enough, and if the magnetic field is large enough,
then this cannot happen.

\subsection{Localization of the Edge Current}

It follows from the calculations done above that the edge current carried
by states $\psi$ of the unperturbed Hamiltonian $H_0$
satisfying $\psi = E_0 ( \Delta_n ) \psi$ are
localized within a region of size $\mathcal{O} ( B^{-1/2})$ near the edge
$x=0$. This corresponds to the classical cyclotron radius.
This is made precise in the following theorem.

\begin{theorem}
Let $\psi$ be a normalized
edge-current carrying state, i.e. $\psi = E_0( \Delta_n )
\psi$, with $\| \psi \| = 1$.
We assume that
the interval $\Delta_n$ as in (\ref{interval1})
satisfies $| \Delta_n | / B$ small,
and that $\mathcal{V}_0 > (2n+3)B$, as in Theorem 2.3.
Then, for any level $n$, any real number $\alpha >- 1 \slash 2$, and for
any $\beta > 0$, there
exist constants $B_{n,\alpha, \beta} > 0$, $C_{n,\alpha, \beta} > 0$, and
$K_{n,\alpha, \beta} > 0$, independent of $B$, so that for
$B > B_{n,\alpha, \beta}$, we have
\beq
\label{local1}
\int_\R ~dy ~\int_{-B^{-\beta}}^{B^{\alpha} } ~dx ~| \psi (x,y) |^2 \geq (1
- C_{n,\alpha, \beta} e^{- K_{n,\alpha, \beta} B^{2 \alpha+1} } ),
\eeq
provided $\mathcal{V}_0 \geq (2n+c)B + B^{2(2 \alpha + \beta + 1)}$.
\end{theorem}

\vspace{.1in}
\noindent
{\bf Proof.} Let $I_{\alpha, \beta}$ be the interval $[ - B^{-\beta},
B^\alpha ]$. To prove (\ref{local1}), we need to show that
\beq
\label{local2}
\int_\R ~dy ~\int_{\R \backslash I_{\alpha, \beta}} ~dx ~| \psi (x,y) |^2
\leq C_{n,\alpha, \beta} e^{- K_{n,\alpha, \beta} B^{2 \alpha+1} } .
\eeq
In light of the expansion (\ref{decomp1})--(2.5), and the normalization $\|
\psi \| = 1$, the integral on the left of (\ref{local2}) has the form
\beq
\label{local3}
\int_\R ~dy ~\int_{\R \backslash I_{\alpha, \beta}} ~dx ~| \psi (x,y) |^2 =
\sum_{j=1}^n \int_{\omega_j^{-1} (\Delta_n)} ~dk | \beta_j (k)|^2 ~\int_{\R
\backslash I_{\alpha, \beta}} ~dx | \varphi_j (x;k)|^2 .
\eeq
Hence, it suffices to prove that the following integrals
\beq
\label{interror1}
\int_{- \infty}^{-B^{-\beta}} \varphi_j (x;k)^2 ~dx, ~\mbox{for} ~k \in
\omega_j^{-1} (\Delta_n),
\eeq
and
\beq
\label{interror2}
\int_{B^{\alpha}}^{+\infty} \varphi_j (x;k)^2 ~dx, ~\mbox{for} ~k \in
\omega_j^{-1} (\Delta_n),
\eeq
are bounded above as in (\ref{local2}).\\

\noindent
{\bf Step 1.} We start by proving that given $\delta > 1 \slash
2$, there is $\tilde{B}_{n,\delta}>0$ such that:
\beq
\label{spe1}
\forall B >\tilde{B}_{n,\delta},\ \forall j=0,1,\ldots,n,\
\omega_j^{-1}(\Delta_n) \subset (-\infty, B^{\delta}).
\eeq
To see this, we consider a $\Cm[2]{\R}$ function $J$ satisfying $J(x)=0$
for $x \leq 0$, and $J(x) =1$ for
$x \geq 1 \slash \sqrt{B}$. Furthermore, we assume that
$\| J' \|_{\infty} \leq C_1
\sqrt{B}$, and $\| J'' \|_{\infty} \leq C_2 B$, for two finite
constants $C_1 , C_2 > 0$.
Let $\psi_n (x;k)$ be the harmonic oscillator eigenfunction given in
(\ref{harmoscef1}). The function $J \psi_n(.;k)$, $k \in \R$,
obviously belongs to the domain of $h_0(k)$. An
easy computation gives
\begin{eqnarray*}
(h_0(k)-(2n+1)B) J(x) \psi_n(x;k) & = & [ h_0(k), J ] \psi_n(x;k) \\
                                  & = & -2i J'(x) \psi_n'(x;k) - J''(x)
\psi_n(x ;k).
\end{eqnarray*}
As the support of $J'$ is contained in $[0, 1 \slash \sqrt{B}]$, we
have the following estimate
\bea
& & \| (h_0(k)-(2n+1)B) J \psi_n(.;k) \| \nonumber\\
& \leq & 2 C_1 \sqrt{B} \| \chi_B \psi_n'(.;k) \| +
C_2 B \| \chi_B \psi_n(.;k) \|, \label{local4}
\eea
where $\chi_B$ is the characteristic function of
$[0, 1 \slash \sqrt{B}]$.
Now, any given $k \geq B^{\delta}$, the explicit expression
(\ref{harmoscef1}) of $\psi_n(.;k)$ assures us
there exist three constants $B'_{n,\delta}>0$, $C'_{n,\delta}>0$ and
$K'_{n,\delta}>0$ such that
\[ \| \chi_B \psi_n(.;k) \| +
\| \chi_B \psi_n'(.;k) \|
\leq C'_{n,\delta} \expo{-K'_{n,\delta} B^{2 \delta -1}}, \]
for any $B > B'_{n,\delta}$.
Inserting this estimate in (\ref{local4}), we immediately
see that $|\omega_n(k)-(2n+1)B|$, can be made smaller than $(a-1)B$ by taking $B$
sufficiently large.  This proves (\ref{spe1}).\\

\noindent
{\bf Step 2.} Any given $\gamma >- 1 \slash 2$, $j=0,1,\ldots,n$
and $k \in \omega_j^{-1}(\Delta_n)$, we compute now a pointwise Gaussian
upper
bound for $\varphi_j(.;k)$ in $[B^{\gamma},+\infty)$.
The eigenfunction $\varphi_j(.;k)$ being normalized, we necessarily have
\[ \int_{1 \slash \sqrt{B}}^{B^{\gamma} \slash 2} \varphi_j^2(x;k) dx \leq
1, \] so there is some $x_+ \in ( B^{-1/2}, B^{\gamma} \slash 2)$ such
that
\beq
\label{expdec1}
\varphi_j(x_+;k) \leq \left( \frac{B^\gamma}{2} - B^{-1/2}
\right)^{-1 \slash 2}  \leq 2 B^{1 \slash 4}.
\eeq
Next, we pick $\delta$ in $(1 \slash 2,\gamma+1)$ such that
$B^{\delta-1} < x_+ \slash 2$ holds for $B$ sufficiently large. Hence,
taking account of (\ref{spe1}),
we can find some $B''_{n,\gamma}>0$ such that,
\[ x^* = \frac{k + \sqrt{(2n+3)B}}{B} < x_+, \]
for $B > B''_{n,\gamma}$. We consequently have
\bea
W_j(x;k) & = & (Bx-k)^2-\omega_j(k) \nonumber \\
         & \geq & (Bx-k)^2- (2n+3)B \nonumber \\
         & \geq & B^2 (x-x^*)((x+x_*)-2k \slash B) \nonumber \\
         & \geq & B^2(x-x_+)^2 >0, \nonumber
\eea
and $W_j'(x;k)=2B^2(x-k \slash B) >0$, for any $x >x_+$, so
Proposition \ref{boundedBelow} in Appendix 2 implies
\beq
\varphi_j(x;k)  \leq  \varphi_j(x_+;k)  \expo{-B \slash 2 (x-x_+)^2},\
\forall x \geq x_+,
\eeq
since $\varphi_j(.;k)$ is solution of the
Schr\"odinger equation $\varphi_j''(x;k)  = W_j(x;k) \varphi_j(x;k)$ in
$(0,+\infty)$. This, together with (\ref{expdec1}) and
the basic inequality $x_+ < B^{\gamma}$ imply
\beq
\label{expdec2}
\forall x \geq B^{\gamma},\ \varphi_j(x;k) \leq 2 B^{1 \slash 4} \expo{-B
\slash 2 (x-B^{\gamma})^2},\
\eeq
provided $B>B''_{n,\gamma}$.\\

\noindent
{\bf Step 3.} Now, for any $\alpha > 1 \slash 2$, we set $\gamma
= (\alpha + 1 \slash 2) \slash 2$ and insert
(\ref{expdec2}) in the integral (\ref{interror2}): For $B >
B''_{n,\gamma}$, we have
\[ \int_{B^{\alpha}}^{+\infty}  \varphi_j(x;k)^2 dx \leq  P_{n,\alpha}(B)
\expo{-B \slash 2 (B^{\alpha}-B^{\gamma})^2}, \]
where $P_{n,\alpha}(B)$ is a polynomial function of $B$.
There are also three constants $B_{n,\alpha}>0$, $C_{n,\alpha}>0$  and
$K_{n,\alpha} \in (0,1)$ and such that
\beq
\label{estintdestra}
\int_{B^{\alpha}}^{+\infty}  \varphi_j(x;k)^2 dx \leq C_{n,\alpha}
\expo{-K_{n,\alpha} B^{2\alpha+1}},
\eeq
provided $B > B_{n,\alpha}$.\\
\noindent
{\bf Step 4.} Any given $j=0,1,\ldots,n$ and $k \in
\omega_j^{-1}(\Delta_n)$,
we turn now to estimating (\ref{interror1}) for some fixed
$\beta>0$. First, the basic inequality
\[ \int_{-B^{-\beta} \slash 2}^0 \varphi_j(x;k)^2 dx \leq 1, \]
assures us there is $x_{-} \in (-B^{-\beta} \slash 2,0)$ such that
\beq
\label{expdec3}
\varphi_j(x_-;k) \leq \sqrt{2} B^{\beta \slash 2}.
\eeq
Then, we choose $\mathcal{V}_0 > (2n+c)B$ so
\[ W_j(x;k)=(Bx-k)^2+V_0(x)-\omega_j(k) \geq \mathcal{V}_0-(2n+c)B >0, \]
for any $k \in \omega_j^{-1}(\Delta_n)$ and $x<0$. Hence, $\varphi_j(.;k)$
being solution of the Schr\"odinger equation
$\varphi_j''(x;k)  = W_j(x;k) \varphi_j(x;k)$, Proposition
\ref{boundedBelow} in Appendix 2 implies,
\[ \forall x \leq x_-,\ \varphi_j(x;k) \leq \varphi_j(x_-;k)
\expo{\sqrt{\mathcal{V}_0-(2n+c)B} (x-x_-)}. \]
Since $x_- \geq - B^{-\beta} \slash 2$, the previous inequality
together with (\ref{expdec3}) lead to
\[    \varphi_j(x;k) \leq \sqrt{2} B^{\beta \slash 2}
\expo{\sqrt{\mathcal{V}_0- (2n+c)B} (x+B^{-\beta} \slash 2)}, \]
for all $x \leq -B^{-\beta}$, so we immediately get:
\[ \int_{-\infty}^{-B^{-\beta}} \varphi_j(x;k)^2 dx \leq
   \frac{B^{\beta}}{\sqrt{\mathcal{V}_0-(2n+c)B}}
\expo{-\sqrt{\mathcal{V}_0-(2n +c)B} B^{-\beta}}. \]
Hence, for any $\alpha > - 1 \slash 2$ and $B \geq 1$, we have
\beq
\label{estintsinistra}
\int_{-\infty}^{-B^{-\beta}}
\varphi_j(x;k)^2 dx \leq  \expo{-B^{2 \alpha +1}},
\eeq
provided $\mathcal{V}_0 > (2n+c)B + B^{2(2 \alpha + \beta +1)}$.
Recalling now that the constant $K_{n,\alpha}$
in (\ref{estintdestra}) is smaller than $1$, the result obviously
follows from (\ref{estintdestra}) and
(\ref{estintsinistra}). $\Box$

\vspace{.1in}
\noindent
We now extend this result to the perturbed case.
We assume that the conditions guaranteeing the
existence of edge current-carrying states for
the perturbed Hamiltonian are satisfied.
In particular, this means that the perturbation
$V_1$ satisfies a bound $\|V_1 \|_\infty < \nu_0 B$,
and that $\tilde{c} - \tilde{a}$ is small enough so
that $\tilde{\Delta}_n$ lies in the spectral gap
of the bulk Hamiltonian $H_{bulk} = H_L (B) + V_\omega$
in the interval $(E_n(B), E_{n+1}(B))$.
We refer the reader to \cite{[CH2],[GK1]}
for a discussion of the properties of $H_{bulk}$.
Under these conditions, the edge current for the perturbed Hamiltonian
remains
close to the wall for all time in a strip of width $B^{- \alpha}$, for any
$\alpha < 1 / 2$, essentially the cyclotron radius.
For any $0 < L_0 < \infty$, we define a spatial truncation function
$0 \leq J_0  \leq 1$ to be $J_0(x) = 0$, for $ x < L_0$ and $J_0 (x) = 1$
for $x > L_0 + 1$.

\vspace{.1in}

\begin{theorem}
Consider the perturbed operator $H = H_0 + V_1$
with $\|V_1 \|_\infty < \nu_0 B$, for some constant $0 < \nu_0 < \infty$.
Let $\Delta_n \subset \tilde{\Delta}_n = [ (2n+\tilde{a})B, (2n+
\tilde{c})B]$
lie in the spectral gap of the bulk Hamiltonian $H_{bulk} =
H_L(B) + V_1$ in $(E_n(B), E_{n+1}(B))$.
Let $\psi = E(\Delta_n) \psi \in L^2 (\R^2)$ be an edge current carrying
state so that the results of Theorem 2.3 hold true. In particular,
we assume that $\nu_0$ and that $\tilde{c} - \tilde{a}$
are small enough so that
the lower bound (\ref{pertcurr1}) is valid.
Then, for any level $n$, and for any $0 < \alpha < 1 \slash 2$,
there exist constants $0 < C_n, K_n < \infty $, independent of $B$,
so that for a strip of width $L_0 = B^{-\alpha}$, we have
\beq
\label{local12}
\| J_0 \psi \| \leq
C_n e^{- K_n  B^{1/2 - \alpha} } .
\eeq
\end{theorem}

\vspace{.1in}
\noindent
{\bf Proof.} The method of proof is similar to that given in \cite{[CHS1]}.
The resolvent formula for $H_{bulk}$ and $H$ gives
\beq
\label{resolv11}
R(z) = R_{bulk} (z) - R_{bulk} (z) V_0 R(z) .
\eeq
Let $0 \leq f \leq 1$
be a smooth, nonnegative function with $f | \Delta_n = 1$
and $\supp f \subset \tilde{\Delta}_n$.
Then, we can write $\psi = f(H) \psi$.
We use the Helffer-Sj\"ostrand
formula for the operator $f(H)$, cf. \cite{[Davies]} or \cite{[CHS1]}.
Let $\tilde{f}$ be an almost analytic extension of $f$ into a small
complex neighborhood of $\tilde{\Delta}_n$ that vanishes of order two as
$\Im z \rightarrow 0$.
The Helffer-Sj\"ostrand formula for $f(H)$ is
\beq
\label{hsj1}
f(H)= \frac{-1}{\pi} \int_\C \partial_{\overline{z}} \tilde{f} (z) (H -
z)^{-1} ~dx ~ dy.
\eeq
Note that since the support of $f$ lies in the spectral gap of
$H_{bulk}$, formula (\ref{hsj1}) shows that $f(H_{bulk}) = 0$.
Then, by the resolvent formula (\ref{resolv11}),
and the Helffer-Sj\"ostrand formula (\ref{hsj1}),
we can write
\bea
\label{geom1}
J_0 \psi &=& J_0 f(H) \psi \nonumber \\
&=&  \frac{-1}{\pi}
      \int_\C \partial_{\overline{z}} \tilde{f} (z) J_0 R_{bulk} (z) V_0 R
(z) ~dx ~dy  .
\eea
The distance between the supports of the confining
potential $V_0$ and the localization function $J_0$ is $0 < L_0 < \infty$.
An application of the Combes-Thomas method to Landau Hamiltonians as
presented, for example, in \cite{[CH2]},
results in the following bound for the operator $J_0 R_{bulk} (z) V_0$ for
$z$ in the resolvent set of $H_{bulk}$. There are constants $0 < C_1, C_2 <
\infty$ so that
\beq
\label{ct1}
\|J_0 R_{bulk} (z) V_0 \| \leq \frac{C_1}{d(\sigma (H_{bulk}), z)} e^{-C_2
B^{1/2} L_0} .
\eeq
The distance $d(\sigma (H_{bulk}), z)$ is given by the minimum of the
distance from the larger interval
$\tilde{\Delta}_n$ to the band edges of the spectrum of $H_{bulk}$ at
$E_n(B) + \| V_1 \|_\infty$ and $E_{n+1} (B) - \| V_1 \|_\infty$.
Consequently, if $L_0 = B^{-\alpha}$, for $\alpha < 1/2$, we obtain the
result.
$\Box$


\renewcommand{\thechapter}{\arabic{chapter}}
\renewcommand{\thesection}{\thechapter}

\setcounter{chapter}{3} \setcounter{equation}{0}

\section{The Straight Edge and Dirichlet Boundary Conditions}

We note that the lower bounds on the edge currents in Theorems 2.1 and 2.3
are independent
of the size of the confining
potential barrier $\mathcal{V}_0$, provided $\mathcal{V}_0 >> E_{n+1} (B)$.
This indicates that these lower bounds should remain valid in the
limit $\mathcal{V}_0 \rightarrow \infty$.
This limit formally corresponds to Dirichlet boundary conditions along
the edge at $x=0$.
In this section, we use the results of section 2.1 and 2.3 to prove
lower bounds on the edge current with Dirichlet boundary conditions (DBC)
along $x = 0$.
DeBi\`evre and Pul\'e \cite{[DBP]}
and Fr\"ohlich, Graf, and Walcher \cite{[FGW]}
both considered the Landau Hamiltonian with
Dirichlet boundary conditions along the edge $x=0$ in their articles.
Both groups proved the existence of edge currents
using the commutator method
described in section 5. We provide an alternate proof of this here.
DeBi\`evre and Pul\'e \cite{[DBP]}
avoid the minor technical difficulty encountered by Fr\"ohlich, Graf, and
Walcher \cite{[FGW]} due to the
nonselfadjointness of $p_x$ on a half line by using $y$ as a conjugate
operator.
We provide an alternate proof of the existence of edge currents in the hard
boundary case here.

We denote the Landau Hamiltonian $H_L(B)$ on the space $L^2 ( [ 0 , \infty
) \times \R)$ with Dirichlet boundary conditions along $x=0$ by $H_0^D$.
This unperturbed operator admits a direct integral
decomposition with respect to the $y$-variable. We denote by $h_0^D(k)$ the
corresponding fibered operator with eigenvalues $\omega_j^D (k)$ and
eigenfunctions $\varphi_j^D(x;k)$.
These eigenfunctions provide an eigenfunction expansion of any
state, as in (\ref{decomp1}), and we denote the coefficients
of this expansion by $\beta_j^D (k)$.
The eigenfunctions of $h_0^D(k)$
are given explicitly by Whittaker functions. Many
properties of the dispersion curves $\omega_j^D(k)$ are derived from the
properties and estimates on Whittaker functions, cf. \cite{[DBP]}.
The perturbed operator is denoted by $H_D
\equiv H_0^D + V_1$, on the same Hilbert space.
We let $E_0^D ( \cdot)$ and $E_D( \cdot)$
denote the corresponding spectral families.
As in section 2, the interval $\Delta_n = [ (2n + a)B, (2n+c)B]$, with
$1<a < c < 3$.

\begin{theorem}
Consider the operators $H_0^D$ and $H_D = H_0^D + V_1$,
on $\mathcal{H} \equiv L^2 ( [ 0 , \infty ) \times \R)$,
with Dirichlet boundary conditions along $x=0$.
Any state $\psi \in E_D ( \Delta_n ) \mathcal{H}$
carries an edge current satisfying the lower bounds
(\ref{current1p}), with the same constants (\ref{ncnst1})--(\ref{ncnst2}),
provided $(c-a)$ and $\| V_1 \|_\infty / B $ are sufficiently small as
stated
there.
\end{theorem}

We prove this theorem through a
perturbation argument comparing $H_0^D$ with $H_0 = H_L(B) + V_0$
in the large $\mathcal{V}_0$ regime.
We begin with an estimate on the trace of the eigenfunctions
$\varphi_j(x;k)$ of $h_0(k)$ on the line $x=0$.

\begin{lemma}
\label{tracelemma1}
Let $\varphi_j (x;k)$ be a normalized
eigenfunction of $h_0(k)$ as in section 2.
For any $0 \leq j \leq n$,
and for all $k \in \omega_j^{-1} (\Delta_n)$, we have
\beq
\label{boundaryest1}
0 \leq \varphi_j (0; k) \leq
\left( \frac{2B }{\mathcal{V}_0} \right)^{1/2}
\; [ (2n+3) B ]^{1/4} .
\eeq
In general, for any eigenfunction $\varphi_l(x;k)$, and for any $k \in \R$,
we have
\beq
\label{boundaryest2}
0 \leq \varphi_l (0;k) \leq \left( \frac{2B}{\mathcal{V}_0} \right)^{1/2}
\omega_l(k)^{1/4} \leq \left( \frac{2B}{\mathcal{V}_0} \right)^{1/2}
[ (2l+3)B + \mathcal{V}_0 ]^{1/4} .
\eeq
\end{lemma}

\noindent
{\bf Proof.}
One can choose $\varphi_j (x;k) \geq 0$, for $x < 0$,
as discussed in Appendix 1, Proposition 8.1.
From Proposition 2.1, and the consequence
of the Feynman-Hellmann
Theorem (\ref{disp1})--(\ref{curr2}), we have
\bea
\label{boundaryest3}
\varphi_j(0;k)^2 & = &
- \frac{2B}{\mathcal{V}_0} \langle \varphi_j (\cdot; k),
\hat{V}_y (k) \varphi_j (\cdot;k)
\rangle \nonumber \\
&=& - \frac{B}{\mathcal{V}_0} \omega_j ' (k) \geq 0,
\eea
as we recall that $\omega_j'(k) \leq 0$. A simple calculation now gives
\bea
\label{slopebound2}
|\omega_j' (k)| &=& | \langle
\varphi_j ( \cdot; k), h_0' (k) \varphi_j (\cdot;k) \rangle| \nonumber \\
&=& 2 | \langle \varphi_j ( \cdot; k), (k-Bx)  \varphi_j (\cdot;k)
\rangle|  \nonumber \\
&\leq& 2 | \langle \varphi_j ( \cdot; k), (k-Bx)^2  \varphi_j (\cdot;k)
\rangle |^{1/2} \nonumber \\
&\leq & 2 \omega_j (k)^{1/2} \leq 2 [ (2n+3)B]^{1/2},
\eea
by positivity of the operator $h_0(k)$, and
the fact that $k \in \omega_j^{-1} (\Delta_n)$.
Combining this with (\ref{boundaryest3}),
we obtain the bound (\ref{boundaryest1}). The bound (\ref{boundaryest2})
follows from (\ref{slopebound2})  and the structure of the dispersion
curves. $\Box$\\

We next show how Lemma \ref{tracelemma1} implies the convergence of the
dispersion curves $\omega_j(k)$ to $\omega_j^D (k)$ as $\mathcal{V}_0
\rightarrow \infty$.
We use an estimate on the eigenvalues $\omega_j^D(k)$
of the Dirichlet problem
that follows from an estimate
in Lemma 2.1 of De Bi\`evre and Pul\'e \cite{[DBP]}. The explicit
properties of the eigenfunctions $\varphi_j(x;k)$ allow one to prove that
if $j \neq l$, then there is  finite a constant $C_{jl} > 0$ so that
\beq
\label{dbpev1}
| \omega_j^D(k) - \omega_l^D(k) | \geq C_{jl} B, ~~ \forall k \in \R.
\eeq

\begin{lemma}
The dispersion curves $\omega_j(k)$ are monotonic increasing functions of
$\mathcal{V}_0$. For $\mathcal{V}_0 >> E_{n+1} (B)$, and for $j = 0 ,
\ldots , n$, and for $k \in \omega_j^{-1}(\Delta_n)$,  we have
\beq
\label{dispconv1}
0 \leq \omega_j^D(k) - \omega_j(k) \leq \frac{ C_0(n,B) }{
\mathcal{V}_0^{1/2}}.
\eeq
\end{lemma}

\vspace{.1in}
\noindent
{\bf Proof.}
The Hamiltonians $h_0(k)$ are analytic operators in the parameter
$\mathcal{V}_0$.
We use the Feynman-Hellmann Theorem to compute the variation of the
eigenvalues
$\omega_j(k)$ with respect to $\mathcal{V}_0$. This gives
\beq
\label{evpos1}
\frac{ \partial \omega_j }{\partial \mathcal{V}_0} (k) = \int_{\R^-}
\varphi_j(x;k)^2 ~dx \geq 0,
\eeq
so that the dispersion curves are monotone increasing with respect to
$\mathcal{V}_0$. Furthermore, the rate of increase in (\ref{evpos1}) slows
as $\mathcal{V}_0 \rightarrow \infty$. This follows from the pointwise upper
bound on $\varphi_j (x,k)$ restricted to $x \leq 0$. In particular,
from (\ref{expupper1}) and the trace estimate (\ref{boundaryest1}), we have
\bea
\label{evbound1}
0 \leq \frac{ \partial \omega_j }{\partial \mathcal{V}_0} (k) &\leq&
\varphi_j (0;k)^2 \int_{-\infty}^0 e^{- 2 \sqrt{ ( \mathcal{V}_0 -
\omega_j(k) ) } |x| } ~dx \nonumber \\
& \leq & \frac{ (2n+3)^{1/2}}{\sqrt{ ( \mathcal{V}_0 -\omega_j(k) ) }}
\left(
\frac{B^{3/2}}{\mathcal{V}_0} \right).
\eea
This shows that the dispersion curve $\omega_j^D(k)$ is an upper bound on
the dispersion curves $\omega_j(k)$.
To prove the rate of convergence (\ref{dispconv1}),
we use the eigenvalue equation
\beq
- \varphi_j '' + (k-Bx)^2 \varphi_j = \omega_j \varphi_j,
\eeq
and take the inner product with the Dirichlet eigenfunction $\varphi_l^D$.
After integration by parts, and an application of the eigenvalue
equation for $\varphi_l^D$, one obtains,
\beq
\label{evdiff1}
( \omega_l^D(k) - \omega_j(k) ) \langle \varphi_l^D (\cdot; k) , \varphi_j
(\cdot; k) \rangle = ( \varphi_l^D )' ( 0 ; k) \varphi_j (0;k) .
\eeq
The estimate in Lemma \ref{tracelemma1} implies that the left side of
(\ref{evdiff1}) vanishes as $\mathcal{V}_0 \rightarrow \infty$,
that is
\beq
\label{evdiff2a}
| \omega_l^D(k) - \omega_j(k) | ~| \langle \varphi_l^D (\cdot; k) ,
\varphi_j(\cdot; k) \rangle | \leq | ( \varphi_l^D )' ( 0 ; k) |
\left( \frac{2B}{\mathcal{V}_0 } \right)^{1/2} [(2n+3)B]^{1/2} .
\eeq
We next show that $|  \langle \varphi_j^D (\cdot; k) , \varphi_j(\cdot; k)
\rangle |$ is uniformly bounded from below as $\mathcal{V}_0 \rightarrow
\infty$, proving the convergence of the eigenvalues. To show this, let
$\chi_\pm$ denote the characteristic functions onto the left and right half
lines $( - \infty , 0]$ and $[0 , \infty)$, respectively. We
first note that
\beq
\label{lbef1}
\| \varphi_j ( \cdot; k) \|^2 = 1 = \| \chi_- \varphi_j ( \cdot; k) \|^2 +
\| \chi_+ \varphi_j ( \cdot; k) \|^2 ,
\eeq
and the upper bound on the eigenfunction $\varphi_j$ on the negative
half-axis (\ref{expupper1}), together with (\ref{boundaryest1}),
imply that
\beq
\| \chi_- \varphi_j ( \cdot; k) \| \leq C_j \mathcal{V}_0^{-3/4},
\eeq
so that
\beq
\label{massl}
\| \chi_+ \varphi_j ( \cdot; k) \| \geq 1 - \mathcal{O}
(\mathcal{V}_0^{-3/4} ) ,
\eeq
as $\mathcal{V}_0 \rightarrow \infty $ and $k \in \omega_j^{-1} (\Delta_n)$.
Now, for $l \neq j$, it follows from (\ref{dbpev1})
and the monotonicity of the dispersion curves in $\mathcal{V}_0$ that
\beq
\label{dbpev2}
|\omega_l^D(k) - \omega_j(k) | \geq | \omega_l^D(k) - \omega_j^D(k) |
\geq C_{lj}  B.
\eeq
So it follows from this (\ref{dbpev2}) and from
(\ref{evdiff2a}) that
for $l \neq j$
\beq
\langle \varphi_l^D (\cdot; k) , \varphi_j(\cdot; k) \rangle
\rightarrow 0, ~~\mbox{as} ~~\mathcal{V}_0 \rightarrow \infty.
\eeq
If, in addition, the matrix element $\langle \varphi_j^D (\cdot; k) ,
\varphi_j(\cdot; k) \rangle$ also vanished as $\mathcal{V}_0 \rightarrow
\infty$, this would contradict (\ref{massl}) as the family $\{ \varphi_l^D
(\cdot;k) \}$ is an orthonormal basis.
It follows that this matrix element must be bounded from below uniformly in
$\mathcal{V}_0$ as $\mathcal{V}_0 \rightarrow \infty$.
Consequently, the dispersion curves must converge
as $\mathcal{V}_0 \rightarrow \infty$ with the specified rate. $\Box$ \\

The local
convergence of the dispersion curves to those for the Dirichlet problem
is a key ingredient in proving the convergence of the projection
$P_j(k)$, for the
eigenvalue $\omega_j(k)$ of $h_0(k)$, to the projector $P_0^D(k)$,
for the eigenvalue $\omega_j^D (k)$ of $h_0^D(k)$, when $\mathcal{V}_0$
tends to infinity (with $B$ fixed). The proof relies on the comparison of
the resolvents $R_0(z;k) = (h_0(k) - z)^{-1}$ and $R_0^D (z;k) = (h_0^D(k)
- z)^{-1}$, as $\mathcal{V}_0 \rightarrow \infty$, for $z \in \Gamma_j
(\mathcal{V}_0)$, a contour of radius $1/ \mathcal{V}_0^{3/8}$ about
$\omega_j^D (k)$, for $0 \leq j \leq n$ and $k \in \Sigma_n$.
The comparison of the resolvents relies on a formula
derived from Green's theorem and various trace estimates. This is rather
standard; we refer, for example, to the discussion in \cite{[HM]}.
This is the content of the next lemma.

\begin{lemma}
Let $P_j(k)$, respectively $P_j^D(k)$, for $j = 0 , \ldots, n$,
be the projection onto the one-dimensional subspace of
$h_0(k)$, respectively $h_0^D(k)$, corresponding to the eigenvalue
$\omega_j(k)$, respectively $\omega_j^D(k)$. Then, there
exists a finite constant $C_1 (n,B) > 0$,
such that for all $\mathcal{V}_0$ sufficiently
large, and uniformly for $k \in
(\omega_j^D)^{-1} ( \Delta_n) \cup \omega_j^{-1} (\Delta_n)$,
we have
\beq
\label{projconv1}
\| P_j (k) - P_j^D (k) \| \leq \frac{ C_1(n,B) }{\mathcal{V}_0^{1/4} }.
\eeq
\end{lemma}

\vspace{.1in}
\noindent
{\bf Proof.}

\noindent
1. Let us recall that the interval $\Delta_n \subset (E_n (B), E_{n+1}
(B))$ is fixed. We are concerned with the first $n+1$-eigenvalues
$\omega_j(k)$ of $h_0(k)$, for $j = 0, \ldots, n$. We fix $0 \leq j \leq
n$, and let $\Gamma_j ( \mathcal{V}_0)$
be the circle of radius $1/ \mathcal{V}_0^{3/8}$
about $\omega_j^D(k)$.
By Lemma 3.2, there is
an amplitude $\mathcal{V}_0^*  >> 1$ so that $| \omega_j^D
(k) - \omega_j (k) | < C_n \mathcal{V}_0^{-1/8}$,
and $\mbox{dist} ~(z,
\omega_j(k) ) \geq \mathcal{V}_0^{-1/4} $,
for $\mathcal{V}_0 > \mathcal{V}_0^*$.
We will always assume this condition. Moreover, there exists
an index $N(\mathcal{V}_0) >> n$, such that if $l > N ( \mathcal{V}_0)$,
we have $\mbox{dist} ~( \omega_l(k), \Gamma_j ( \mathcal{V}_0)
) > \mathcal{V}_0$. The index $N ( \mathcal{V}_0)$ can be chosen to be
proportional to $\mathcal{V}_0$ since $\omega_l(k)$ is bounded above by
$(2l+3)B + \mathcal{V}_0$.
In order to estimate the difference of the projectors on the left
in (\ref{projconv1}), we use the contour representation of the projections
in terms of the resolvents so that
the difference of the projectors is
written as
\beq
\label{approxproj2}
P^D_j (k) - P_j (k) = \frac{1}{2 \pi i} \int_{\Gamma_j( \mathcal{V}_0)}
\; ( R_0 (z;k) - R_0^D (z;k) ) ~dz.
\eeq
The resolvent formula for the difference of the two resolvents
in (\ref{approxproj2}) following from Green's theorem is
\beq
\label{resolv1}
R_0(z;k) - R_0^D(z;k) = R_0(z;k) T_0^* B_0 R_0^D(z;k) ,
\eeq
where $T_0$ is the trace map $(T_0 u) (x) = u(0)$, and $(B_0u)(x) = u'(0)$.
The trace map is a bounded map from $H^1 ( \R )\rightarrow \C$.
Due to the simplicity of the eigenvalues, the resolvent $R_0 (z;k)$
has the expression
\beq
\label{resexp1}
R_0 (z;k) = \sum_{j=0}^\infty \frac{P_j(k)}{\omega_j(k) - z},
\eeq
where $P_j(k)$ projects onto the one-dimensional subspace
spanned by $\varphi_j (x;k)$.
Substituting (\ref{resolv1}) into the right side of
(\ref{approxproj2}), we obtain
\beq
\label{approxproj3}
P^D_j (k) - P_j (k) = \frac{1}{2 \pi i} \int_{\Gamma_j( \mathcal{V}_0)}
R_0 (z;k) T_0^* B_0 R_0^D (z;k) ~dz .
\eeq

\noindent
2. We now estimate the integral of (\ref{approxproj3}) for $z \in \Gamma_j(
\mathcal{V}_0)$ and $k \in  \omega_j^{-1} ( \Delta_n)$.
We decompose any $\phi \in L^2 (\R)$
into a piece $\phi^L$ supported on $( - \infty, 0]$,
and its complement: $\phi = \phi^L + \phi^R$.
With this decomposition applied to any $\phi, \psi \in L^2 (\R)$,
we write the inner product of the difference of the resolvents as
\beq
\label{innerprod1}
\langle \phi, ( R_0(z;k) - R_0^D(z;k) ) \psi \rangle = \langle \phi^R, (
R_0(z;k) - R_0^D(z;k) ) \psi^R \rangle +   \mathcal{E}_{LR} (z;k) .
\eeq
The mixed error term $\mathcal{E}_{LR}$ has the form
\beq
\label{errorLR1}
\mathcal{E}_{LR} (z;k) = \langle \phi^L, R_0(z;k) \psi^R \rangle +
\langle \phi, R_0(z;k) \psi^L \rangle .
\eeq
For the first term of (\ref{innerprod1}),
we have,
\bea
\label{matrixele1}
| \langle \phi^R, ( R_0(z;k) - R_0^D(z;k) ) \psi^R \rangle |
& = & | \langle T_0 R_0 (\overline{z}; k) \phi^R, B_0 R_0^D (z;k) \psi^R
\rangle | \nonumber \\
& \leq & | T_0 R_0 (\overline{z}; k) \phi^R | \| \psi^R \| \| T_0 \|_{H^1,
\C}
\| p_x R_0^D (z;k ) \|_{L^2, H^1} . \nonumber \\
& &
\eea
The trace is evaluated using the expansion (\ref{resexp1}) and the estimate
(\ref{boundaryest1}). Using the Cauchy-Schwarz inequality, we obtain
\bea
\label{tracest2}
| T_0 R_0 (\overline{z}; k) \phi^R | & \leq & \sum_{l \geq 0} \left|
\frac{ \varphi_l(0;k)}{ \omega_l (k) - z } \right| \cdot |
\langle \varphi_l (\cdot; k) , \phi^R \rangle | \nonumber \\
& \leq & \left( \sum_{l \geq 0} \frac{ |\varphi_l(0;k)|^2}{| \omega_l (k)
- z |^2 } \right)^{1/2} \| \phi \| .
\eea
We split the sum into two parts: $l=0, \ldots, n$, and $l > n$, where $n$
is independent of $B$ and $\mathcal{V}_0$.
With $z \in \Gamma_j ( \mathcal{V}_0)$,
and the estimate on the trace $\varphi_j(0;k)$ given in
(\ref{boundaryest1}),
we obtain an estimate for the first sum
\beq
\label{firstsum1}
\left( \frac{2B}{\mathcal{V}_0} \right)^{1/2}
\left( \sum_{j=0}^n \frac{ [ (2n+3)B ]^{1/2} }{
| \omega_j (k) - z |^2 } \right)^{1/2}
\leq \frac{ C_2 (n,B) }{ \mathcal{V}_0^{1/4} },
~~z \in \Gamma_j (\mathcal{V}_0) .
\eeq
For the second sum, we need to use the second estimate
(\ref{boundaryest2}) for the
trace. We again split the sum into two parts: $n < l \leq N(\mathcal{V}_0)
$,
and $l > N( \mathcal{V}_0 ) $. The first sum is finite and easily seen to be
bounded by $C_2 (n,B) \mathcal{V}_0^{- 1/4}$.
The second sum is bounded as
\beq
\label{secondsum2}
\left( \frac{2B}{\mathcal{V}_0} \right)^{1/2}
\left( \sum_{l > N ( \mathcal{V}_0 )   }
\frac{ [ (2l+1)B + \mathcal{V}_0 ]^{1/2} }{ | \omega_l (k) - z |^2 }
\right)^{1/2}
\leq \frac{C_3 (n,B)}{ \mathcal{V}_0^{3 /4}} .
\eeq
Hence, the sum in (\ref{tracest2}) is at worse
order of $\mathcal{V}_0^{-1/4}$.
Returning to the estimate in (\ref{matrixele1}),
it is simple to check that
\beq
\label{sobolev1}
\| p_x R_0^D (z;k ) \|_{L^2, H^1} \leq C_4 \mathcal{V}_0^{3/8}, ~~z \in
\Gamma_j (\mathcal{V}_0) .
\eeq
Combining estimates (\ref{firstsum1})--(\ref{sobolev1}),
and recalling that the length of the contour $\Gamma_j
(\mathcal{V}_0)$ is $\mathcal{V}_0^{-3/8}$,
we find that
\beq
\label{resolvest2}
\left| \int_{ \Gamma_j ( \mathcal{V}_0)}
\langle \phi^R, ( R_0(z;k) - R_0^D(z;k) ) \psi^R \rangle ~dz \right| \leq
\left( \frac{ C_5 (n,B) }{ \mathcal{V}_0^{1/4}} \right) \| \phi \| ~\| \psi
\|.
\eeq

\noindent
3. The error term $\mathcal{E}_{LR}$ in (\ref{errorLR1})
is evaluated by substituting the expansion (\ref{resexp1})
into each inner product of $\mathcal{E}_{LR}$.
We then separate each sum into three sets of indices.
For the first two sets of indices,
$0 \leq j \leq n$, and
$n < j \leq  N(\mathcal{V}_0)$, the half-line $x<0$ is in the
classically forbidden region for the eigenfunctions $\varphi_j (x;k)$, with
$k \in \omega_j^{-1} ( \Delta_n )$.
For the third set of indices, we have
$\mbox{dist} ~(z, \omega_l (k) ) >> C_0 \mathcal{V}_0$.
For the first two sets of indices, that is for $0 \leq l \leq
N(\mathcal{V}_0)$, it follows from section 8 that
the eigenfunctions $\varphi_l(x;k)$ satisfy the bound
\beq
\label{efdecay1}
\varphi_l(x;k) \leq \varphi_l(0;k) e^{- \sqrt{ V_0 - \omega_l (k) } |x|} ,
~~\mbox{for} ~x \leq 0.
\eeq
We begin with the matrix element
\bea
\label{me1}
M_1 & \equiv & | \langle \phi^L , R_0 (z;k) \psi^R \rangle | \nonumber \\
& = & \left|
\sum_{l=0}^\infty \frac{ \langle \phi^L, \varphi_l(\cdot;k) \rangle
\langle \varphi_l(\cdot;k), \psi^R \rangle}{ \omega_l(k) - z} \right|
\nonumber \\
& \leq & M_{1,1} + M_{1,2} + M_{1,3},
\eea
where $M_{1,n}$, for $n = 1,2,3$, denote the sum over the indices in each
of the three sets indicated above.
For the first sum $M_{1,1}$, we use the exponential decay (\ref{efdecay1})
and the Cauchy-Schwarz inequality to obtain an upper bound on the matrix
element
\beq
\label{me2}
| \langle \phi^L ,\varphi_l(\cdot;k) \rangle| \leq \frac{ | \varphi_l (0;k)
| }{[ 4 ( \mathcal{V}_0 - \omega_l (k) ) ]^{1/4} } \| \phi \| .
\eeq
Using the estimate (\ref{boundaryest1})
for the trace, and the fact that $z \in
\Gamma_j (\mathcal{V}_0)$,
the term $M_{1,1}$ is bounded as
\bea
\label{errorLR2}
M_{1,1} &=&  \left| \sum_{j =0}^n
\frac{ \langle \phi^L , \varphi_j ( \cdot;k)
\rangle ~\langle \varphi_j (\cdot; k), \psi^R \rangle}{ \omega_j(k) - z}
    \right| \nonumber \\
& \leq & \left( \sum_{j=0}^n \frac{ | \varphi_j (0;k)
|^2 }{ 2 ( \mathcal{V}_0 - \omega_l (k) )^{1/2} } ~\frac{1}{ | \omega_j(k)
- z|^2} \right)^{1/2} ~\| \phi \| ~\| \psi \| \nonumber \\
& \leq & \frac{C_6(n,B)}{ \mathcal{V}_0^{1/2}} ~\| \phi \| ~\| \psi \|.
\eea
For the second term $M_{1,2}$, we use estimate (\ref{boundaryest2})
for the trace. The relevant sum is
\beq
\label{me3}
\sum_{l=n+1}^{N(\mathcal{V}_0)} \left( \frac{2B}{\mathcal{V}_0} \right)
~\frac{ [ (2l+3)B
+\mathcal{V}_0 ]^{1/2} }{ 2 ( \mathcal{V}_0 - \omega_l(k) )^{1/2} } ~
\frac{1}{| \omega_l(k) -z|^2}
\leq \frac{ C_7(n,B)}{ \mathcal{V}_0}^{1/2} .
\eeq
Finally, for $M_{1,3}$, we use the fact that the term in the denominator of
(\ref{me1}) satisfies $| \omega_l(k) - z| \geq C_0 \mathcal{V}_0 $.
Hence, the estimates (\ref{errorLR1})--(\ref{me3}) are bounded by order
$\mathcal{V}_0^{-1/4}$.
By the same methods, the second matrix element $M_2$ in (\ref{errorLR1})
appearing in $\mathcal{E}_{LR}$ is easily seen to be order of
$\mathcal{V}_0^{-1/4}$.
Returning to the contour integral of the error term $\mathcal{E}_{LR}$,
we see that
\beq
\label{errorLR3}
\left| \int_{\Gamma_j (\mathcal{V}_0)} \mathcal{E}_{LR} (z;k) ~dz \right|
\leq \frac{C_8(n,B)}{\mathcal{V}_0^{5/8}} \| \phi \| ~\| \psi \|.
\eeq
This estimate, and the estimate (\ref{resolvest2})
of the main term prove the result
(\ref{projconv1}). $\Box$

\vspace{.1in}

\noindent
{\bf Proof of Theorem 3.1.}
We begin with the unperturbed case.
Let $\psi \in L^2 ( \R^+ \times \R)$ satisfy $\psi =
E_0^D ( \Delta_n ) \psi$. We assume that the hypotheses of Lemma 2.1 hold
so that there are no
cross-terms in the matrix element $\langle \psi, V_y \psi \rangle$.
We will use the results of Lemma 3.2 that tell us that $\omega_j (k)
\rightarrow \omega_j^D (k)$, locally,
and that the matrix element $\langle \varphi_j^D
( \cdot; k ), \varphi (\cdot ; k) \rangle \geq D_0$, as $\mathcal{V}_0
\rightarrow \infty$.
We write
\bea
\label{unpertvel1}
- \langle \psi, V_y \psi \rangle &=&
- \sum_{j=0}^n \int_{ (\omega_j^D)^{-1} (\Delta_n) } ~dk
  | \beta_j^D (k)|^2 \langle
\varphi_j^D ( \cdot; k), P_j^D(k)
\hat{V}_y (k) P_j^D (k) \varphi_j^D ( \cdot; k) \rangle  \nonumber \\
  &\geq & - \sum_{j=0}^n \int_{(\omega_j^D)^{-1} (\Delta_n) }
   ~dk | \beta_j^D (k)|^2 | \langle \varphi_j^D ( \cdot; k), \varphi_j
(\cdot;
k) \rangle |^2 \nonumber \\
& & \times  \langle \varphi_j ( \cdot; k), P_j(k) \hat{V}_y (k) P_j (k)
  \varphi_j ( \cdot; k) \rangle  - \mathcal{R} (\psi) \nonumber \\
&\geq & - D_0 \sum_{j=0}^n \int_{(\omega_j^D)^{-1} (\Delta_n) }
  ~dk | \beta_j^D (k)|^2
  \langle \varphi_j ( \cdot; k), P_j(k) \hat{V}_y (k) P_j (k)
  \varphi_j ( \cdot; k) \rangle   \nonumber \\
& &  - \mathcal{R} (\psi) .
\eea
The remainder $\mathcal{R} (\psi)$ is bounded by
\beq
\mathcal{R} (\psi)
\leq  2 \sum_{j=0}^n \int_{( \omega_j^D) ^{-1} (\Delta_n) }
   ~dk  | \beta_j^D (k)|^2
      \left\{ | \langle \varphi_j^D ( \cdot; k), ( P_j^D(k) - P_j(k) )
      \hat{V}_y (k) P_j^D (k) \varphi_j^D ( \cdot; k) \rangle | \right\} .
\eeq
The main term in (\ref{unpertvel1}) is bounded from below as in Theorem
2.1. Estimates on the difference of the spectral projectors given
in Lemma 3.3
establish the appropriate bounds on the remainder $\mathcal{R} (\psi)$.
This proves the theorem for the unperturbed case.
The perturbation theory of section 2.2 now
applies in the same manner as in that section. $\Box$


\renewcommand{\thechapter}{\arabic{chapter}}
\renewcommand{\thesection}{\thechapter}

\setcounter{chapter}{4} \setcounter{equation}{0}

\section{One-Edge Geometries with More General Boundaries}

The previous
results were based on the exact calculations for the unperturbed case due
to the possibility of taking the partial Fourier transform.
Fr\"ohlich, Graf, and Walcher \cite{[FGW]} considered more general one-edge
geometries
for which the boundary satisfies some mild regularity conditions.
We first review these results, and then present some new results
based on the notion of the
{\it asymptotic velocity of edge currents} coming from scattering theory.
These results apply to a very general class of perturbations of the
half-plane geometry.

Fr\"ohlich, Graf, and Walcher \cite{[FGW]} studied one-edge, simply
connected,
unbounded regions $\Omega \subset \R^2$,
with a piecewise $C^3$-boundary. The boundary must satisfy some additional
geometric conditions
so that the edge does not asymptotically become parallel to itself so that
the region resembles a two-edge geometry near infinity. If this occurs,
the interaction of the classical
trajectories in different directions may cancel each other.
The authors consider the
unperturbed Hamiltonian $H_0^D$ which is the Landau Hamiltonian on $\Omega$
with Dirichlet boundary conditions on $\partial \Omega$.
The main theorem of \cite{[FGW]} is the following.

\begin{theorem}
Assume that the region $\Omega$
satisfies the geometric conditions discussed above and that the
perturbation
$V_1 \in L^\infty ( \R^2)$.
Let $E/B \notin 2 \N + 1$ and suppose
that $B$ is taken sufficiently large so
that $\| V_1 \|_\infty / B$ is sufficiently small.
Then, the spectrum of $H_\Omega^D
= H_0^D + V_1$ is absolutely continuous
near $E$.
\end{theorem}

As in the work of DeBi\`evre and Pul\'e \cite{[DBP]},
and as we discuss in section 5, Fr\"ohlich, Graf, and Walcher
construct a conjugate operator for the Hamiltonian $H_\Omega^D$
on the region $\Omega$.
They prove that the commutator, when spectrally
localized to a small interval of energies
around $E$, has a strictly positive lower bound.
Mourre theory \cite{[CFKS]}
then implies the existence of absolutely continuous spectrum near $E$.
The Dirichlet boundary conditions on $\partial \Omega$ cause
some technical complications as $p_x$ is not self-adjoint on any domain.
The conjugate operator is a quantization of a linearization of
the classical guiding center trajectory for the
classical electron orbit.

We introduce another notion to the study of geometrically perturbed regions
and use it to prove the persistence of edge currents.
The {\it asymptotic velocity} is defined for any pair of
self-adjoint Schr\"odinger operators $(H_0, H)$ for which the wave
operators exist. The (global)
wave operators $\Omega_\pm$ for the pair $(H_0 , H)$ are defined by
\beq
\label{waveop1}
\Omega_\pm \equiv s-\lim_{t \rightarrow \pm \infty} e^{i t H} e^{-itH_0}
E_{ac} (H_0) ,
\eeq
where $E_{ac} (H_0)$ is the projection onto the absolutely continuous
spectral subspace for $H_0$. When the wave operators exist, the range is
contained in the absolutely continuous spectral subspace of $H$, and
the wave operators are partial isometries between these spectral subspaces.
We will use the local wave operators
$\Omega_\pm (\Delta)$ obtained by replacing $E_{ac} (H_0)$ by the projector
$E_0(\Delta)$
for $H_0$ and
an interval $\Delta$ in the absolutely continuous subspace of $H_0$.
The asymptotic velocity is defined for any component of the velocity
observable. We
are interested in velocity asymptotically in the $y$-direction and for
states
with energy in an interval $\Delta$.
We define this to be
\beq
\label{asympvel2}
V_y^{\pm} (\Delta) \equiv \Omega_\pm (\Delta) V_y \Omega_\pm^* (\Delta).
\eeq
We note that when $H_0$ commutes with $V_y$, and the local wave operators
exist, the local
asymptotic velocity is obtained by the limit
\beq
\label{asympvel122}
V_y^{\pm} (\Delta) \equiv s-\lim_{t \rightarrow \pm \infty} e^{itH} E_0
(\Delta)V_y E_0(\Delta) e^{-itH}.
\eeq
In the context of potential scattering, we refer to the book of Derezinski
and G\'erard \cite{[DG]} for a complete discussion of the {\it asymptotic
velocity}.

We consider the geometric perturbation of the
straight, one-edge geometry obtained by
perturbing the boundary confining potential
$V_0$. We recall that a {\it sharp confining potential} $V_0$ is
a constant multiple $\mathcal{V}_0 >> 0$ of the
characteristic function $\chi_\Omega$ for a region $\Omega$.
In section 2, we treated the case
$\Omega = \Omega_0 \equiv ( - \infty, 0] \times \R$, the half-plane.
Here, we consider more general
$\Omega$ obtained by perturbing the half-plane $\Omega_0$.

\vspace{.1in}
\noindent
{\bf Condition C.} {\it The sharp confining potential $V_\Omega$
is supported in a region $\Omega$ so that $\Omega \backslash \Omega_0$
lies in the strip $|y| \leq R < \infty$, for some $0 < R < \infty$.}

\vspace{.1in}
We first consider the pair of Hamiltonians $(H_0, H_\Omega)$,
where $H_0= H_L(B) + V_0$ is the straight-edge Hamiltonian with sharp
confining
potential, and $H_\Omega = H_L(B) + V_\Omega$, describes the geometric
perturbation of the straight-edge
boundary satisfying Condition C.  We prove that the local
wave operators exist for this pair and that the asymptotic velocity
observable is bounded from below by $B^{1/2}$. This observable corresponds
to the edge current at $y = \pm \infty$. Furthermore, the spectrum
of the perturbed operator $H_\Omega$ still has absolutely continuous
spectrum
between the Landau levels. We then show that this lower bound
on the asymptotic velocity observable
is stable under a perturbation $V_1$ that is small compared to the
field strength $B$.

\vspace{.1in}
\begin{theorem}
\label{asymptvelth0}
Let $H \equiv H_L(B) + V_\Omega + V_1$
be the perturbed
Hamiltonian with sharp confining potential $V_\Omega$
and a bounded perturbation $V_1 \in
L^\infty ( \R^2)$. Suppose the region $\Omega \backslash \Omega_0$
satisfies Condition $C$.
Let $\Delta_n$
be an energy interval between Landau levels as in (\ref{interval1}).
Let $V_y^\pm (\Delta_n)$
be the asymptotic velocity for the pair $(H_0, H_\Omega)$. Suppose
that $(c-a)$ and $\| V_1 \|_\infty / B$ are sufficiently small as in
Theorem 2.3.
For any state $\psi = E (\Delta_n) \psi$,
the asymptotic edge-current
velocity $V_y^\pm (\Delta_n)$ satisfies
\beq
\label{asympvel0}
\langle \psi, V_y^\pm (\Delta_n) \psi \rangle \geq  C_n B^{1/2} \| \psi
\|^2.
\eeq
\end{theorem}

We remark that is is not required that the new region $\Omega$
be connected nor that it be bounded in the $x$-direction.
The basic situation that we have in mind, however, is the one for which
the new region $\Omega$ represents a distortion of the
boundary of the half-plane $\Omega_0$.
It is interesting to note that the edge
current persists for some states even if the boundary extends to $+ \infty$
along the $x$-axis.
For example, the right half-plane may actually be disconnected if the
perturbation is supported in a
cone-type region with vertex at $y=0$ and $x = + \infty$.

Before we prove Theorem \ref{asymptvelth0}, we consider the effect of
the boundary perturbation with $V_1 = 0$.
We define $H_0 = H_L(B) + V_0$ and $H_\Omega = H_L(B) + V_\Omega$, and
we denote the corresponding spectral families by $E_0 (\cdot)$ and $E_\Omega
(\cdot)$, respectively.
We first prove the existence of the
local wave operators for the pair $(H_0, H_\Omega)$ by the method of
stationary phase.
This proves the existence of absolutely continuous spectrum in intervals
between Landau levels. We then use these
local wave operators to prove the persistence of edge currents.
We consider the perturbation of the confining potential $V_0 (x)$ given by
\beq
\label{pertpot1}
V_\Omega (x,y) = \mathcal{V}_0
( \chi_{(-\infty, 0]} (x) + \chi_{\Omega \backslash \Omega_0} (x,y))
= V_0 (x) + \mathcal{V}_0 \chi_{\Omega \backslash \Omega_0 } (x,y) ,
\eeq
and we will write $\delta V \equiv V_\Omega - V_0$,
so that $\delta V = \mathcal{V}_0 \chi_{\Omega \backslash \Omega_0 }
(x,y)$. This perturbation of the confining potential is interpreted as a
perturbation of the boundary of the region where the electron can
propagate.

\begin{proposition}
\label{waveop3}
Let $\Delta_n$ be as in (\ref{interval1})
with $(c-a)$ sufficiently small.
Then, the local wave operators $\Omega_\pm (\Delta_n)$
for the pair $(H_0, H_\Omega)$ exist. As a consequence, operator $H_\Omega$
has absolutely continuous spectrum in $\Delta_n$.
\end{proposition}

\noindent
{\bf Proof.}
We use Cook's method and study the local operators defined by
\bea
\label{waveop4}
\Omega (t; \Delta_n) - E_0 (\Delta_n) & =  &
e^{i t H_\Omega } e^{-i H_0 t} E_0 (\Delta_n) - E_0 (\Delta_n) \nonumber \\
&=& i \int_0^t ~ e^{i s H_\Omega } \delta V e^{-i H_0 s}
     E_0 (\Delta_n) ~ds .
\eea
Hence, it suffices to prove that for any smooth vector $\psi$,
\beq
\label{cauchy1}
\lim_{t_1, t_2 \rightarrow \infty}
\int_{t_1}^{t_2}\delta V e^{-i s H_0 } E_0 (\Delta_n)\psi ~ds  = 0.
\eeq
In order to prove (\ref{cauchy1}), we use the method of stationary phase.
Using the partial Fourier transform
in (\ref{cauchy1}), we have
\bea
\label{cauchy2}
\lefteqn{ ( \delta V e^{-i s H_0 } E_0 (\Delta_n)\psi ) (x,y)} \nonumber \\
&=&  \sum_{j = 0}^n \delta V (x,y)
\int_\R e^{-i \omega_j (k) s + iky}
\chi_{\omega_j^{-1} (\Delta_n)} (k) \hat{\psi} (x,k) ~dk .
\eea
We define the phase as $\Phi (k,y,s) \equiv ky - \omega_j (k) s$,
and note that the derivative is
$\partial_k \Phi (k,y,s) = y - \omega_j' (k) s$.
Let $\chi_R (y)$ be the characteristic function on the interval $[-R, R]$.
We have the following lower bound
\beq
\label{cauchy3}
| \partial_k \Phi (k,y,s) \chi_{\omega_j^{-1} (\Delta_n)} (k) \chi_R (y) |
\geq | \omega_j ' (k)s - y | \chi_{\omega_j^{-1} (\Delta_n)} (k) \chi_R (y)
.
\eeq
In section 2.2, we proved that
\bea
\label{cauchy4}
- \omega_j'(k) \chi_{ \omega_j^{-1} (\Delta_n ) } (k) & \geq &
\frac{\mathcal{V}_0}{2B} \varphi_j (0;k)^2
\chi_{ \omega_j^{-1} (\Delta_n ) } (k) \nonumber \\
&\geq & \frac{(\mathcal{V}_0 - \omega_j (k) )}{2B^3 \mathcal{V}_0 C_n}
( \omega_j (K) - E_n(B))^2 (E_{n+1}(B) - \omega_j(k))^2
\chi_{ \omega_j^{-1}(\Delta_n) } (k) \nonumber \\
  & \geq & C_{n,j} B  \chi_{\omega_j^{-1}(\Delta_n)} (k) .
  \eea
Using this lower bound
(\ref{cauchy4}) in the lower bound (\ref{cauchy3}), we obtain
\beq
\label{cauchy5}
| \partial_k \Phi (k,y,s) \chi_{\omega_j^{-1} (\Delta_n)} (k) \chi_R (y) |
\geq (C_{n,j}Bs - R)\chi_{\omega_j^{-1} (\Delta_n)} (k) \chi_R (y) .
\eeq
As a consequence,
we can differentiate the phase factor in (\ref{cauchy1}) and bound
the integral there by
\beq
\label{cauchy6}
\sum_{j = 0}^n \frac{1}{\langle s \rangle^N}
\left|  \int_{\omega_j^{-1} ( \Delta_n )}
( \partial_k^N e^{i \Phi (k,y,s)} ) \hat{\psi}(x,k) ~dk \right| ,
\eeq
where $\langle s \rangle \equiv ( 1 + |s|^2 )^{1/2}$.
The convergence of the integral in (\ref{cauchy1}) follows from this
decay and integration by parts using the smoothness of $\psi$. $\Box$

\vspace{.1in}

\begin{proposition}
Assume the hypotheses of Proposition \ref{waveop3}.
For any $\psi \in E_\Omega (\Delta_n) L^2 (\R^2)$, we have
\beq
\label{asympvel1}
\langle \psi , V_y^\pm (\Delta_n) \psi \rangle \geq  C_n B^{1/2} \| \psi \|^2,
\eeq
where the constant $C_n$ is as in Theorem 2.1.
That is, the asymptotic velocity $V_y^\pm ( \Delta_n)$ of the
edge current carried by the state $\psi = E_\Omega (\Delta_n) \psi$,
for the perturbed region,
is bounded from below by $B^{1/2}$.
\end{proposition}

\noindent
{\bf Proof.}
As a consequence of the existence of the wave operators,
we have the local intertwining relation
\beq
\label{inter1}
\Omega_\pm (\Delta_n)^* E_\Omega (\Delta_n) \psi
=  E_0 (\Delta_n)  \Omega_\pm (\Delta_n)^* \psi.
\eeq
This intertwining property (\ref{inter1}) and the definition
(\ref{asympvel2}) show that
\bea
\label{asympvel23}
\langle \psi, V_y^\pm (\Delta_n ) \psi \rangle &=& \langle \psi,
\Omega_\pm (\Delta_n) E_0 (\Delta_n)
V_y E_0 (\Delta_n) \Omega_\pm^* (\Delta_n) \psi \rangle \nonumber \\
&=& \langle E_0 (\Delta_n) \Omega_\pm (\Delta_n) \psi,
    V_y E_0 (\Delta_n) \Omega_\pm (\Delta_n) \psi  \rangle .
\eea
The lower bound for the right side of (\ref{asympvel23})
follows from Theorem 2.1,
\bea
\label{asympvel3}
\langle E_0 (\Delta_n) \Omega_\pm (\Delta_n) \psi, V_y E_0 (\Delta_n)
\Omega_\pm (\Delta_n) \psi
&  \geq &  C_n B^{1/2} \| E_0 (\Delta_n) \Omega_\pm (\Delta_n) \psi  \|^2
\nonumber \\
& \geq & C_n B^{1/2} \| \Omega_\pm^* (\Delta_n) \psi  \|^2  .
\eea
Since the wave operators are
partial isometries, we have the normalization
\beq
\| \psi \| = \| \Omega_\pm^* ( \Delta_n)  \psi \| ,
\eeq
which, together with (\ref{asympvel3}),
proves the lower bound in (\ref{asympvel1}). $\Box$

\vspace{.1in}
\noindent
We now prove the stability of the edge current
with respect to a small perturbation $V_1 \in L^\infty ( \R^2)$.
Although we do not
necessarily know the spectral type of the
perturbed Hamiltonian in intervals between the Landau levels,
the edge current is stable.

\vspace{.1in}
\noindent
{\bf Proof of Theorem 4.2.}
The proof of Theorem 4.2 follows the same
lines of the proof of Theorem 2.3. Given
$\psi$ as in the theorem, we decompose
it according to the spectral projectors for $H$
and a slightly larger
interval $\tilde{\Delta}_n$ containing $\Delta_n$. As in (\ref{decomp1p}),
we
write
\beq
\label{boundarypert1}
\psi = E_\Omega(\tilde{ \Delta}_n) \psi +
E_\Omega (\tilde{\Delta}_n^c) \psi \equiv \phi + \xi.
\eeq
We then have the decomposition as in (\ref{ident1}). We bound $\| \xi \|$ as
in (\ref{xibound1}), and in order
to bound $\| V_y^\pm (\Delta_n) \xi \|$, we note that the asymptotic
velocity is bounded
by definition
\beq
\label{boundaqrypert2}
\| V_y^\pm ( \Delta_n ) \| \leq [( 2n+c) B ]^{1/2} ,
\eeq
as follows from (\ref{asympvel122}).
Finally, we note that the matrix element for $\phi$ satisfies
\beq
\label{boundarypert3}
\langle \phi, V_y^\pm (\Delta_n) \phi \rangle \geq \tilde{C}_n B^{1/2} \|
\phi \|^2,
\eeq
by Proposition 4.2. A simple calculation as in the proof of Theorem 2.3
allows us to obtain the lower bound
\beq
\label{boundarypert4}
\| \phi \|^2 \geq \left[ 1 - \left( \frac{c-a}{\tilde{c} - \tilde{a}} +
\frac{2 \| V_1 \| }{ B ( \tilde{c} -
\tilde{a} ) } \right) \right] \| \psi \|^2,
\eeq
so by taking $c-a$ and $\| V_1 \| / B$ sufficiently small, we obtain the
result (\ref{asympvel0}). $\Box$


\renewcommand{\thechapter}{\arabic{chapter}}
\renewcommand{\thesection}{\thechapter}

\setcounter{chapter}{5} \setcounter{equation}{0}

\section{One-Edge Geometries and the Spectral Properties of $H= H_0 + V_1$}

The unperturbed operator $H_0 = H_L(B) + V_0$
has purely absolutely spectrum and $\sigma (H_0) = [B , \infty)$.
In the paper \cite{[DBP]},
DeBi\`evre and Pul\'e proved
that perturbations $V_1$, as in Theorem 2.3, preserve
the absolutely continuous spectrum
in an interval $\Delta_n$, provided $|\Delta_n| = c-a$ is sufficiently
small.
We mention this result here for completeness, and for
comparison with the situation for two-edge geometries where we will use
commutator methods. For a review of commutator methods, we refer the reader
to \cite{[ABMG], [CFKS], [Mou]}.
The proof in \cite{[DBP]} relies on the commutator identity
\beq
\label{mourre1p}
i[ H_0 , y ] = 2 V_y .
\eeq
This commutator shows that an estimate
on the edge current
is equivalent to an estimate on the positivity of the commutator.
This, in turn, provides an estimate on the spectral type of $H_0$.
As we will see,
this equivalence, that an estimate on the edge current implies a commutator
estimate, no
longer holds for two-edge and other, more complicated geometries.
This is one of the reasons we presented a different
approach to the one-edge
geometries in the previous sections.

Continuing with the perturbation theory of $H_0$,
the commutator on the left in (\ref{mourre1p})
is invariant under any perturbation of $H_0$ by a real-valued
potential provided $V_1$ and $y$ have a common, dense domain. It follows
immediately from the commutator
\beq
\label{mourre2p}
i [ H_0 + V_1 , y ] = 2 V_y,
\eeq
and the techniques of Theorem 2.3,
that if $c-a$ is small enough, there exists a finite constant
$K_n > 0$ such that
\beq
\label{mourreest1}
E ( \Delta_n ) (i[ H , y ] ) E ( \Delta_n ) \geq K_n E( \Delta_n).
\eeq
Since the double commutator is $[[H, y], y] = -2i$, the following theorem
now
follows from standard Mourre theory (cf. \cite{[CFKS]}).

\vspace{.1in}
\begin{theorem}
Let $V_1$ satisfy the conditions of Theorem 2.3. If $c-a$ and $\| V_1
\|_\infty / B$ satisfy the
smallness conditions of Theorem 2.3 with respect to $n$ and $B$, then the
operator $H = H_0 + V_1$ has only absolutely continuous spectrum on
$\Delta_n$.
\end{theorem}

Thus, in
the half-plane case, the existence of edge currents for each $\psi \in
E(\Delta_n) L^2 (\R^2)$ is equivalent to the
existence of absolutely continuous spectrum.
This is need not be the case, however, for more complicated edge geometries.
For those situations, there may be edge currents carried by states $\psi$
but the spectrum need not be absolutely continuous (cf. \cite{[HS2],
[EJK], [FM0],[FM1],
[FM2]}).


\renewcommand{\thechapter}{\arabic{chapter}}
\renewcommand{\thesection}{\thechapter}

\setcounter{chapter}{6} \setcounter{equation}{0}

\section{One-Edge Geometries and General
Confining Potentials}

We prove that the analysis used in section 2 can be extended to the
case of more general confining potentials with a straight edge. These
potentials are described as {\it soft} potentials, as opposed to the {\it
hard} potentials such as the Sharp Confining Potential or Dirichlet boundary
conditions. In general, the soft confining potential $V_0$, supported on $x
\leq 0$, should be rapidly increasing for $x < 0$.
There are two classes of soft confining potentials that we can treat: 1)
{\it convex-concave potentials} that are initially convex and
then become asymptotically flat, such
as $\mathcal{V}_0 \tanh B |x|$, for $x \leq 0$,
and 2) {\it globally convex potentials}, such as monomials $|x|^p$, for $x
\leq 0$ and $p \geq 1$.
These two classes of soft confining
potentials require slightly different hypotheses in order to obtain upper
and lower pointwise exponential bounds on the eigenfunctions of $h_0(k)$.

We make the following assumption on $V_0 \in
\Hploc{(-\infty,x_\varepsilon)}$,
where the point $x_\epsilon$ is defined in (\ref{turningpt2}).
Assumption (H1) is common for both classes of soft confining potentials and
describes the behavior near the turning point.

\begin{tabular}{ll}
& \\
{\bf (H1)} & There is $x_{\varepsilon}$ satisfying (\ref{turningpt2}) for
some
$\varepsilon \in (0,1]$, such that,\\
& $0 \leq V_0(t) \leq (2n + c + 2 / \varepsilon )B \leq V_0(x)$, for all $x
\leq
x_\varepsilon \leq t$.\\
& \\
\end{tabular}\\

\noindent
Moreover, we impose on $V_0$ one of the two following conditions. For soft
confining potentials of type 1 (convex-concave), we require

\begin{tabular}{ll}
& \\
{\bf (H2)} & $|V_0'(t)| \leq 5 B^{3/2} \slash \sqrt{2 \varepsilon}$
for a.e. $t < x_{\varepsilon}$. \\
\end{tabular}

\vspace{.1in}
\noindent
For soft confining potentials of type 2 (monomial), we require

\begin{tabular}{cl}
& \\
{\bf (H2')} & For any $k \in \Sigma_n$, there is $C_k>0$ such that the
double
inequality,\\
& $-C_k \sqrt{(Bt-k)^2+V_0(t)-(2n+c)B} \leq V_0'(t) + 2B(Bt-k) \leq 0 $,\\
& holds for a.e. $t < x_{\varepsilon}$.\\
&
\end{tabular}\\
Roughly speaking,
condition (H2') means
that the confining potential
$V_0$ lies in between two parabolas in $(- \infty, x_\varepsilon)$.
We note that the size of the potential depends on the energy level one is
studying. The constant $C > 0$ in (H2), for example, depends on the Landau
level $n$.

Concerning soft confining potentials of type 1, we note that many examples
can
be constructed satisfying hypotheses (H1) and (H2).
The proto-type soft confining potential
of type 1
is a deformation of the Sharp Confining Potential. Indeed, we show that
the Sharp Confining Potential, given by
\beq
V_0(x)=\mathcal{V}_0 \chi_{(-\infty,x_\varepsilon)}(x),
\eeq
with
$\mathcal{V}_0 \geq (2n+c+2 \slash \varepsilon) B$,
and treated in section 2 with $\varepsilon = 0$ by
other methods, satisfies conditions (H1) and
(H2).
Among soft confining potentials obtained as deformations
of the Sharp Confining Potential, we note the
exponential potential
\beq
\label{expconf1}
V_0(x)  =  \mathcal{V}_0 \chi_{(-\infty,0)}(x) ( e^{\alpha / B^{1/2} |x| }
- 1),
\eeq
for $\mathcal{V}_0 \sim \mathcal{O}(B)$ sufficiently large, depending on
$n$, and a constant
$\alpha > 0$.
Other examples can be constructed from the hyperbolic tangent
$\mathcal{V}_0 \tanh B^{1/2} |x|$, for $x \leq 0$, and the inverse tangent
$\mathcal{V}_0 \tan^{-1} B^{1/2} |x|$, for $x \leq 0$.

Our primary example
of a soft confining potential $V_0$ of type 2
is the Parabolic Confining Potential given by
\beq
\label{parpot1}
V_0(x) =  \mathcal{V}_0 \chi_{(-\infty,0)}(x) x^2,
\eeq
with $\mathcal{V}_0 \geq (2n+c+2 \slash \varepsilon) B \slash \sqrt{(-
x_\varepsilon)}$. We will verify in Lemma \ref{lm-Wpara} of Appendix 3
that this Parabolic Confining Potential (\ref{parpot1})
satisfies (H1) and (H2').

We can also treat soft confining potentials given by decreasing monomials
of the form
\beq
\label{parpotp}
V_0(x)  = \mathcal{V}_0 \chi_{(-\infty,0)}(x) (-x)^p,
\eeq
for $p > 1$ and $\mathcal{V}_0$ sufficiently large. The method
required for these confining potentials is slightly different.
Moreover, these confining potentials are of interest in the two-edge
geometries treated in paper 2 \cite{[HS2]}. For these reasons,
the proofs are given there.

For the unperturbed model $H_0= H_L + V_0$, we have the following result.

\begin{theorem}
\label{thm-gal}
Let $V_0$ be a confining potential on the half-plane satisfying (H1)
together with (H2) (resp. (H2')).
Then, for any $\psi = E_0 ( \Delta_n ) \psi$ having an expansion as in
(\ref{decomp1})
with coefficients $\beta_j (k)$, there is a constant $C_{n,\varepsilon} > 0$
so that for all $| \Delta_n| / B$ small enough, we have
\beq
\label{gencurr1}
- \langle \psi , V_y \psi \rangle \geq C_{n,\varepsilon} (a-1)^2(3-c)^2
\left(
  \sum_{j=0}^n \int_{\omega_j^{-1} (\Delta_n)} ~dk | \beta_j (k) |^2
\left( \frac{\tilde{V}_j(k)}{V_j(k)^2} \right) \right) B^{1/2},
\eeq
where $\tilde{V}_{j,\varepsilon}(k)$ is defined by (\ref{const5}) (resp. by
(\ref{const5bis})) and
$V_{j,\varepsilon}(k)$ by (\ref{const6}) (resp. by (\ref{const6bis})).
\end{theorem}

\noindent
{\bf Proof.} We prove the statement for $V_0$ satisfying (H1) and (H2).
We also assume that the conditions of Lemma 2.1 are satisfied so that the
cross-terms vanish.
We begin with the formula for the matrix element $\langle \psi , V_y
\psi \rangle$ in (\ref{gencurr1}) following from the partial Fourier
transform,
\bea
\label{gencurr2}
- \langle \psi , V_y \psi \rangle &=& \frac{-1}{2B} \sum_{j=0}^n
  \int_{-\infty}^0 ~dx \int_{\omega_j^{-1} (\Delta_n)} ~dk | \beta_j (k)
|^2 \varphi_j (x;k)^2 V_0'(x) \nonumber \\
&\geq & \frac{-1}{2B} \sum_{j=0}^n \int_{-\infty}^{x_\varepsilon} ~dx
\int_{\omega_j^{-1} (\Delta_n)} ~dk | \beta_j (k)
|^2 \varphi_j (x;k)^2 V_0'(x) .
\eea
The strategy is to use the lower bound of
Proposition \ref{prop-sfl}
for $| \varphi_j (x;k)|$ and to obtain a lower bound for $| \varphi_j
(x_\varepsilon; k)| $. We first turn to estimating
$| \varphi_j(x_\varepsilon; k)| $.
We use the results of Lemma 2.2. We expand the eigenfunctions
$\varphi_j (x;k)$ in terms of the harmonic oscillator eigenfunctions
$\psi_m(x;k)$ given in (\ref{harmoscef1}), as in (\ref{expand2}).
We find that
\beq
\label{coefest1}
\sum_{m=0}^n | \alpha_m^{(j)} (k) |^2 \geq \frac{1}{2B(n+1)} (E_{n+1} (B) -
\omega_j (k) ),
\eeq
and, with $P_n$ denoting the projector onto the subspace of $L^2 (\R)$
spanned by the first $n$ harmonic oscillator eigenfunctions,
\beq
\label{matrix1}
| \langle \varphi_j ( \cdot, k ), V_0 P_n \varphi_j ( \cdot, k) \rangle |
\geq \frac{1}{2B(n+1)} (\omega_j (k) - E_n(B)) (E_{n+1}(B) - \omega_j(k)).
\eeq
We also need an upper bound on this matrix element (\ref{matrix1}).
From the definition of $P_n$, we obtain
\beq
\label{matrix2}
| \langle \varphi_j ( \cdot, k ), V_0 P_n \varphi_j ( \cdot, k) \rangle |
\leq \sum_{m=0}^n | \alpha_m^{(j)} (k)| \{ I_{j,m}(x_\varepsilon;k) +
II_{j,m} (x_\varepsilon;k) \}
\eeq
where the integrals $I_{j,m}$ and $II_{j,m}$ are given by
\beq
\label{defI}
I_{j,m}(x_\varepsilon;k) \equiv \int_{-\infty}^{x_\varepsilon} V_0(x) ~|
\varphi_j(x;k)| ~| \psi_m (x;k)| ~dx,
\eeq
and
\beq
\label{defII}
II_{j,m}(x_\varepsilon;k) \equiv \int_{x_\varepsilon}^0 V_0(x) ~|
\varphi_j(x;k)| ~| \psi_m (x;k)| ~dx .
\eeq
We estimate (\ref{defII}) using hypothesis (H1),
\beq
\label{potest1}
0 \leq V_0 (x) \leq ( 2/ \varepsilon + 2n + c) B, ~~\mbox{for}
~x_\varepsilon
< x \leq 0,
\eeq
and the form of the harmonic oscillator wavefunction (\ref{harmoscef1}),
giving
\begin{equation}
\label{intII1}
II_{j,m}(x_\varepsilon;k) \leq ( 2/ \varepsilon + 2n + c)
\frac{B^{5/4}}{\pi^{\quartp}
} \frac{1}{\sqrt{2^m m!} }
\mathcal{H}_{m,\varepsilon} (k) ~|x_\varepsilon|^{1/2} ,
\end{equation}
where the constant $\mathcal{H}_{m,\varepsilon}$ is defined by
\beq
\label{const1}
\mathcal{H}_{m,\varepsilon} (k) \equiv \sup_{x_\varepsilon \leq x \leq 0} |
H_m
(x \sqrt{B} - k / \sqrt{B} ) | e^{- B/2( x - k / B)^2}.
\eeq
The first integral $I_{j,m}$ is estimated as
\beq
\label{intI1}
I_{j,m}(x_\varepsilon;k)\leq  \left( \frac{B}{\pi} \right)^{1/4}
\frac{\tilde{\mathcal{H}}_{m,\varepsilon} (k)}{\sqrt{2^m m!}}
~\int_{-\infty}^{x_\varepsilon} V_0 (x) | \varphi_j
(x;k)| ~dx
\eeq
where the constant $\tilde{\mathcal{H}}_{m,\varepsilon} (k)$ is defined by
\beq
\label{const2}
\tilde{\mathcal{H}}_{m,\varepsilon} (k) \equiv \sup_{x \leq x_\varepsilon}
|  H_m (x \sqrt{B} - k / \sqrt{B} ) | e^{- B/2( x - k / B)^2} .
\eeq
We return to (\ref{matrix2}). In light of the lower bound on the matrix
element given in (\ref{matrix1}) and the upper bounds on the integrals
given in (\ref{intII1}) and (\ref{intI1}), we solve for the integral
in (\ref{intI1}). First, note that an application of the Cauchy-Schwarz
inequality to the sums over $m$ in (\ref{intII1})--(\ref{intI1}) yields
\beq
\label{const3}
\sum_{m=0}^n | \alpha_m^{(j)} (k) |
\frac{\mathcal{H}_{m,\varepsilon} (k)}{\sqrt{2^m m!}} \leq
\mathcal{H}_{n,\varepsilon}^{(j)}(k),
\eeq
and, similarly,
\beq
\label{const4}
\sum_{m=0}^n | \alpha_m^{(j)} (k) |
\frac{\tilde{\mathcal{H}}_{m,\varepsilon}(k)}{ \sqrt{2^m m!}}
\leq \tilde{\mathcal{H}}_{n,\varepsilon}^{(j)}(k) .
\eeq
We obtain the lower bound
\bea
& & \int_{-\infty}^{x_\varepsilon} V_0 (x) | \varphi_j(x;k)| ~dx
\label{lowerbd1}\\
& \geq & \frac{1}{2 \tilde{\mathcal{H}}_{n,\varepsilon}^{(j)} (k) }
\left( \frac{ (\omega_j (k) - E_n(B))
( E_{n+1} (B) - \omega_j (k) ) }{ 2B(n+1)} \right) \left( \frac{\pi}{B}
\right)^{1/4},
\eea
provided the turning point $x_\varepsilon$ satisfies the bound
\beq
\label{turningpt2}
- x_\varepsilon < \left( \frac{(a-1) (c-3 )}{4 (n+1)
\mathcal{H}_{n,\varepsilon} (2/ \varepsilon + 2n +c) } \right)^2
\left( \frac{\pi}{B} \right)^{1/2},
\eeq
where
\beq
\label{defHnt}
\mathcal{H}_{n,\varepsilon} = \max_{j=0,\ldots,n} \sup \{
\mathcal{H}_{n,\varepsilon}^{(j)}(k),\ k \in \omega_j^{-1}(\Delta_n) \}
\eeq
Note that the right side of the bound in (\ref{turningpt2}) is
$\mathcal{O}(B^{- 1/2})$.
We can now estimate $|\varphi_j (x_\varepsilon;k)|$ using this bound and the
pointwise upper bound on $\varphi_j$ in the classically forbidden region
and proved in Proposition \ref{prop-sfl} of Appendix 3,
\beq
\label{uptwbd1}
| \varphi_j(x;k)| \leq | \varphi_j (x_\varepsilon;k)| e^{-\sqrt{2 \slash
\varepsilon B } (x_\varepsilon-x)} ,\ \forall x \leq x_\varepsilon,
\eeq
since the potential $W_j(t;k) \equiv (Bt-k)^2 + V_0 (t) - \omega_j
(k) \geq 2 \slash \varepsilon B$ for any $k \in \omega_j^{-1}(\Delta_n)$.
In light of this upper bound, we define a function $V_{j,\varepsilon}(k)$ by
\beq
\label{const5}
V_{j,\varepsilon} (k) \equiv \int_{-\infty}^{x_\varepsilon} V_0 (x)
e^{-\sqrt{2
\slash \varepsilon B}(x_\varepsilon- x)} dx \geq 0.
\eeq
We insert (\ref{uptwbd1})
into the integral in (\ref{lowerbd1}), rearrange, and obtain
\bea
\label{lowerbd2}
& & |\varphi_j (x_\varepsilon;k)| \\
&  \geq  & \frac{1}{V_{j,\varepsilon}(k)}
\left( \frac{1}{2 \tilde{\mathcal{H}}_{n,\varepsilon}^{(j)} (k) } \frac{
(\omega_j (k) - E_n(B))
( E_{n+1} (B) - \omega_j (k) ) }{ 2B(n+1)} \left( \frac{\pi}{B}
\right)^{1/4} \right). \nonumber
\eea
We return to the expression for the matrix element
of the edge current (\ref{gencurr2}). We use the lower bound on the
eigenfunction $\varphi_j (x;k)$ derived in Proposition \ref{prop-sfl} of
Appendix 3 :
\beq
\label{lptwbd1}
| \varphi_j(x;k)| \geq | \varphi_j (x_\varepsilon;k)| e^{-
(1+ \varepsilon) \int_{x}^{x_\varepsilon} \sqrt{W_j (t;k)} dt },\ \forall x
\leq
x_\varepsilon.
\eeq
We substitute this expression (\ref{lptwbd1}) into the right side of
(\ref{gencurr2}). It will be convenient to introduce another constant
$\tilde{V}_{j,\varepsilon} (k)$ defined by
\beq
\label{const6}
\tilde{V}_{j,\varepsilon} (k) \equiv - \int_{-\infty}^{x_\varepsilon} V_0'
(x)
e^{-2 ( 1+ \varepsilon) \int_{x}^{x_\varepsilon} \sqrt{W_j (t;k)} dt } ~dx
\geq 0 .
\eeq
Notice that $(H2)$ implies that both integrals $V_{j,\varepsilon}(k)$ and
$\tilde{V}_{j,\varepsilon} (k)$ converge.
Next, using the estimate (\ref{lowerbd2}), we obtain
\beq
\label{gencurr3}
- \langle \psi , V_y \psi \rangle  \geq
C_{n,\varepsilon} (a-1)^2(3-c)^2\left( \sum_{j=0}^n \int_{\omega_j^{-1}
(\Delta_n) } ~dk | \beta_j (k) |^2
\left( \frac{\tilde{V}_{j,\varepsilon}(k)}{V_{j,\varepsilon}(k)^2} \right)
\right) B^{ \demip},
\eeq
where
\beq
\label{defCneps}
C_{n,\varepsilon}=\frac{\pi^{\demip}}{2^5 (n+1)^2
\tilde{\mathcal{H}}_{n,\varepsilon}^2},
\eeq
and
\beq
\label{deftiHn}
\tilde{\mathcal{H}}_{n,\varepsilon}= \max_{j = 0 \ldots n} \sup \{
\tilde{\mathcal{H}}_{n,\varepsilon}^{(j)}(k),\ k \in \omega_j^{-1}(\Delta_n)
\}.
\eeq
Now, in the case where $V_0$ satisfies (H2') instead of (H2), it
suffices to notice that the method remains
valid if we substitute
\beq
\label{const5bis}
V_{j,\varepsilon}(k)= \int_{-\infty}^{x_\varepsilon} V_0 (x)
e^{-\int_{x}^{x_\varepsilon} \sqrt{W_j(t;k)} dt} dx,
\eeq
for (\ref{const5}), and
\beq
\label{const6bis}
\tilde{V}_{j,\varepsilon} (k)= - \int_{-\infty}^{x_\varepsilon} V_0' (x)
e^{-2
( 1+ C_k \slash B \varepsilon) \int_{x}^{x_\varepsilon} \sqrt{W_j(t;k)} dt }
~dx,
\eeq
for (\ref{const6}).
$\Box$\\

We now consider the perturbation of $H_0$ by a bounded potential
$V_1(x,y)$. As in Section 2.3,
we consider a larger interval
\[ \tilde{\Delta}_n = [ (2n + \tilde{a})B, (2n+\tilde{c})B], ~~\mbox{for}
~1 < \tilde{a} < a < c < \tilde{c} < 3, \]
containing $\Delta_n$,
and with the same midpoint $(2n + (a+c) \slash 2)B \in  \Delta_n$, and
prove that the edge current survives if
$\| V_1 \|_\infty$ is sufficiently small relative to $B$.

\vspace{.1in}
\noindent
\begin{theorem}
Let $V_0$ satisfy assumptions (H1) and (H2) (resp. (H2')).
Let $V_1 (x,y)$ denote a bounded potential
and $E (\Delta_n)$ be the spectral projection for $H = H_0 + V_1$ and
the interval $\Delta_n$. Let
$\psi \in L^2 (\R^2)$ be a state satisfying $\psi = E (
\Delta_n ) \psi$, and the following condition. Let $\phi \equiv E_0 (
\tilde{\Delta}_n ) \psi$ have an expansion as in (\ref{decomp1})
with coefficients $\beta_j (k)$ satisfying
\beq
\label{Epscoefass}
\Sum_{j=0}^n \int_{\omega_j^{-1} ( \Delta_n )} | \beta_j(k)|^2 \left(
\frac{\tilde{V}_{j,\varepsilon}(k)}{V_{j,\varepsilon}^2(k)} \right) dk \geq
(1/2) \| \phi \|^2,
\eeq
where $V_{j,\varepsilon}(k)$ and $\tilde{V}_{j,\varepsilon}(k)$ are defined
by
(\ref{const5})-(\ref{const6}) (resp. by
(\ref{const5bis})-(\ref{const6bis})).
Then, we have,
\beq
- \langle \psi, V_y \psi \rangle \geq
B^{1/2} ~( ( C_{n,\varepsilon} \slash 2) ( 3 - \tilde{c})^2 (\tilde{a} -
1)^2
- F_{\varepsilon}( n, \| V_1 \| / B ) ) ~\| \psi \|^2,
\eeq
where $C_{n,\varepsilon}$ is defined in (\ref{defCneps}) and
\begin{eqnarray*}
F_\varepsilon(n, \| V_1 \| / B ) & = &
\left( \frac{2}{ ( \tilde{c} -\tilde{a})} \right)^{1/2} \left(
\frac{(c-a)}{2} +
\frac{ \|V_1 \|}{B} \right)^{1/2} \left( 2n + c +
\frac{ \|V_1 \|}{B} \right)^{1/2}  \\
& & \times \left[ 2 + \left( \frac{2}{( \tilde{c} -\tilde{a})} \right)
\left(\frac{(c-a)}{2} +
\frac{ \|V_1 \|}{B} \right) \right] \\
& & + \frac{C_{n,\varepsilon}}{2}  \left( \frac{2}{ ( \tilde{c} -\tilde{a})}
\right)^2 \left( \frac{(c-a)}{2} +
\frac{ \|V_1 \|}{B} \right)^2  ( 3 - \tilde{c} )^2 (\tilde{a} - 1)^2  .\\
& &
\end{eqnarray*}
\end{theorem}

\vspace{.1in}
\noindent
{\bf Proof.}
As in the Proof of Theorem 2.1, we first decompose the function $\psi$ as
\beq
\psi = E_0 (\tilde{\Delta}_n) \psi + E_0 (\tilde{\Delta}_n^c) \psi
\equiv \phi + \xi,
\eeq
and obtain immediately,
\beq
\langle \psi, V_y \psi \rangle = \langle \phi, V_y \phi \rangle + 2 Re
\langle \phi, V_y \xi \rangle + \langle \xi, V_y \xi \rangle.
\eeq
Next, we use (\ref{vxibound2}) and (\ref{vxibound3}) to bound
$|2 Re \langle \phi, V_y \xi \rangle| + \langle \xi, V_y \xi \rangle|$, and
deduce from Theorem \ref{thm-gal} and (\ref{Epscoefass}) :
\begin{eqnarray*}
-\langle \phi, V_y \phi \rangle & \geq &
C_{n,\varepsilon} (a-1)^2(3-c)^2 \left( \sum_{j=0}^n \int_{\omega_j^{-1}
(\Delta_n) } ~dk | \beta_j (k) |^2
\left( \frac{\tilde{V}_{j,\varepsilon}(k)}{V_{j,\varepsilon}(k)^2} \right)
\right) B^{\demip} \\
& \geq & ( C_{n,\varepsilon} \slash 2) (a-1)^2(3-c)^2 B^{1 \slash 2} \|
\phi \|^2.
\end{eqnarray*}
Now, inserting (\ref{xibound1}) in the identity
\beq
\| \phi \|^2=\|\psi \|^2 - \| \xi \|^2,
\eeq
we get
\beq
\| \phi \|^2 \geq \left( 1 - \left( \frac{2}{ ( \tilde{c} -\tilde{a})}
\right)^2 \left(\frac{(c-a)}{2} + \frac{ \|V_1 \|}{B} \right)^2 \right)
\| \psi \|^2,
\eeq
so the result follows by elementary computations. $\Box$

\renewcommand{\thechapter}{\arabic{chapter}}
\renewcommand{\thesection}{\thechapter}

\setcounter{chapter}{7} \setcounter{equation}{0}

\section{Appendix 1: Basic Properties of Eigenfunctions and Eigenvalues of
$h_0(k)$}

After reducing the operator $H_0 = - \Delta + V_0$
to the operator $h_0(k)$ on $L^2 (\R)$ due to the $y$-translational
invariance, we are concerned with studying
the properties of $h_0(k)$ defined by
\beq
\label{reducedop1}
h_0(k)=p_x^2 + (Bx-k)^2+V_0(x) = p_x^2 + V(x;k),
\eeq
where $p_x^2 = - d^2 / dx^2$,
and the nonnegative potential $V_0 (x) \in L^2_{loc} (\R)$.
The resolvent of the operator $h_0(k)=p_x^2 + V(x;k)$ is
compact since the effective potential $V(x;k)=(Bx-k)^2+V_0(x)$ is
unbounded as $|x| \rightarrow \infty$, so
the spectrum is discrete with only $\infty$ as an accumulation point.
We denote the eigenvalues of $h_0(k)$ in increasing order
and denote them by $\omega_j(k)$, $j \geq 0$. The
normalized eigenfunction associated to $\omega_j(k)$ is
$\varphi_j(x;k)$.
The variational method shows that the domain of $h_0(k)$ is
\beq
\label{domain}
~\mbox{dom}(h_0(k)) = \{ \psi \in \Hp{\R} \cap \Lp{\R; w(x;k) dx},\ (p_x^2
+ V( .;k)) \psi \in \Lp{\R} \},
\eeq
with $w(x;k)=(1 + V(x;k))^{\demip}$.
It is a subset of $\Hploc[2]{\R}$
since the effective potential $V(.;k) \in \Lploc{\R}$. We first discuss the
regularity properties of the eigenfunctions. The Sobolev
embedding theorem states that
$\Hploc[2]{\R} \subset \Cm[1]{\R}$, and we have the following property of
the eigenfunctions.

\begin{proposition}
\label{lm-regular}
The eigenfunctions of $h_0(k)$, given by $\varphi_j(.;k)$, are
continuously differentiable in $\R$ for any $j \in \N$
and $k \in \R$. Furthermore, an eigenfunction $\varphi_j(.;k) \in
\Cm[n+2]{I}$
for any open subinterval $I$ of $\R$ such that $V_0 \in \Cm[n]{I}$, $n \geq
0$.
\end{proposition}
\noindent
{\bf Proof.}
The proof of this proposition follows from the Sobolev Embedding Theorem
which gives $\Hploc[2]{\R} \subset \Cm[1]{\R}$, and the fact that the
Schr\"odinger equation
\[
\varphi_j''(x;k) = (V(x;k)-\omega_j(k)) \varphi_j(x;k),
\]
shows that $\varphi_j''(x;k) \in \Lploc{\R}$. $\Box$ \\

In the particular case of the Sharp Confining Potential
$V_0(x)=\mathcal{V}_0 \chi_{(-\infty,0)}(x)$, Proposition \ref{lm-regular}
shows that
$\varphi_j(.;k) \in \Cm[1]{\R} \cap \Cm[\infty]{\R \backslash \{ 0 \} }$.
Notice that
$\varphi_j(.  ;k)$ is continuously differentiable at the origin although
$V_0$ is discontinuous at this point. For the Parabolic Confining Potential
$V_0(x)=\mathcal{V}_0 x^2
\chi_{(-\infty,0)}(x)$,
we have $\varphi_j(.;k) \in \Cm[3]{\R} \cap \Cm[\infty]{\R^*}$ since $V_0$
is only $C^1$ in any neighborhood of the origin.

We next turn to a proof of the simplicity of
the eigenvalues of $h_0(k)$. We state Lemma \ref{lm-UCT}
without proof. It is a simple consequence of the
Unique Continuation Theorem for Schr\"odinger Operators
(Theorem XIII.63 of \cite{[ReedSimon4]}). We will use this lemma in
the proof of
Propositions \ref{lm-simplicity} and \ref{prop-negative}.

\begin{lemma}
\label{lm-UCT}
Let $I$ be an open (not necessarily bounded) subinterval of $\R$, $W \in
\Lploc{I}$ and $\psi \in \Hploc[2]{I}$ satisfy
\[ \psi''(x) = W(x) \psi(x),\ a.e.\ x \in I. \]
Then, if $\psi$ vanishes in the neighborhood of a single point $x_0 \in I$,
$\psi$ is identically zero in $I$.
\end{lemma}

\begin{proposition}
\label{lm-simplicity}
The eigenvalues $\omega_j (k)$ of the operator $h_0 (k)$ are simple for all
$k \in \R$.
\end{proposition}

\noindent
{\bf Proof.} We consider two $L^2$-eigenfunctions $\varphi$ and $\psi$ of
$h_0(k)$ with same energy $E$.
As follows from Proposition \ref{lm-regular}, they are both
$\Hploc[2]{\R}$-solutions of the Schr\"odinger equation
\begin{equation}
\label{eq-sim}
u''(x)=  (V(x,k)-E) u(x),\ a.e.\ x \in \R.
\end{equation}
By substituting $\varphi$ (resp. $\psi$) for $u$ in (\ref{eq-sim}),
multiplying by $\psi$ (resp. $\varphi$),
and taking the difference of the two equalities, we get
\[ \varphi'' (x) \psi (x)  - \varphi (x) \psi''(x)=
   (\varphi' \psi - \varphi \psi' )'(x)= 0,\ a.e.\ x \in \R.
\]
Consequently, the function $(\varphi' \psi - \varphi \psi')$ is a constant
for a.e. $x$ in $\R$, and
this constant is zero since the function is in $L^2(\R)$ as follows from
Proposition \ref{lm-regular},
\beq
\label{crossed}
(\varphi'  \psi - \varphi \psi')(x) = 0,\ \forall x \in \R.
\eeq
Now we notice there is always a real number $a$ such that the
potential $V(x;k)-E >0$ for a.e. $x>a$ (since
$V(x;k) \rightarrow \infty$ as $x \rightarrow \infty$) and $\psi(a) \neq 0$
($\psi$ would be identically zero in $\R$ by Lemma \ref{lm-UCT} otherwise)
so
$\psi(x) \neq 0$ for any $x > a$ by part 1 of Proposition
\ref{prop-negative}. Hence (\ref{crossed}) implies
\[ \left( \varphi \slash \psi \right)'(x)=0,\ \forall x >a,   \]
so we have $\varphi=\lambda \psi$ on $(a,+\infty)$ for some constant
$\lambda \in \R$.
The function $\varphi - \lambda \psi$ is also an $\Hploc[2]{\R}$-solution
to (\ref{eq-sim}) which vanishes in $(a,+\infty)$. It
is also identically zero in $\R$ by Lemma \ref{lm-UCT} hence
$\{ \varphi, \psi \}$ is a one dimensional manifold of $\Lp{\R}$. $\Box$

\renewcommand{\thechapter}{\arabic{chapter}}
\renewcommand{\thesection}{\thechapter}

\setcounter{chapter}{8} \setcounter{equation}{0}

\section{Appendix 2: Pointwise Upper and Lower Exponential Bounds on
Solutions to Certain ODEs}

We obtain pointwise, exponential, upper and lower bounds on solutions to the
ordinary differential equation $\psi'' = W \psi$, with $W > 0$. We apply
these results in the next section to the eigenfunctions $\varphi_j(.;k)$
of $h_0(k)$ in the
classically forbidden region where $W_j(x;k) \equiv V(x;k) -
\omega_j(k) > 0$.
We consider the following general situation.
We let $\psi$ denote a {\it real}  $\Hp{(-\infty,a)}$-solution to the
system
\begin{equation}
\label{systeme}
\left\{ \begin{array}{l}
              \psi''(x)=W(x) \psi(x),\ a.e.\ x < a\\
              \dsp{\lim_{x \rightarrow a^-} \psi(x)=
\psi(a)>0},
          \end{array} \right.
\end{equation}
for some $a \in \R$, where $W \in \Lploc{(-\infty,a)}$ is such that :
\beq
\label{positivePot}
W(x)>0,\ a.e.\ x < a.
\eeq
Standard arguments already used in the proof of Proposition
\ref{lm-regular}, assure us that the solution $\psi \in
\Hploc[2]{(-\infty,a)}$ so $\psi \in \Cm[1]{(-\infty,a)}$.
Moreover $\psi$ is left continuous at $a$, according to (\ref{systeme}).

\subsection{Basic Properties of $\psi$}

We prove the following basic result that characterizes the behavior of the
solution $\psi$ in the classically forbidden region where $W(x) > 0$.

\begin{proposition}
\label{prop-negative}
Any real $\Hp{(-\infty,a)}$-solution $\psi$ to (\ref{systeme}) satisfies :
\begin{enumerate}
\item $\psi(x)>0$ and $\psi'(x)>0$, for any $x < a$;
\item $\dsp{\lim_{x \rightarrow -\infty} W(x) \psi^2(x) = 0}$.
\end{enumerate}
\end{proposition}

\noindent
We prove the first part of Proposition \ref{prop-negative}
in two elementary lemmas.

\begin{lemma}
\label{lm1}
Under the hypotheses of Proposition \ref{prop-negative},
suppose that $\psi(x_0) \psi'(x_0)<0$, for some $x_0 < a$.
If $\psi (x_0 ) > 0$, we have $\psi(x) >\psi(x_0)$, for any $x < x_0$, and
if $\psi (x_0) < 0$, we have $\psi (x_0) > \psi (x)$, for any $x < x_0$.
Consequently, we have $\psi(x) \psi'(x) \geq 0$, for any $x < a$.
\end{lemma}

\vspace{.1in}
\noindent
{\bf Proof.}
We assume that $\psi(x_0)>0$ so that the hypothesis implies
that $\psi'(x_0)<0$. The case $\psi(x_0)<0$, implying
$\psi '(x_0)>0$, is treated in the same manner.
Notice that $\mathcal{E}=\{ \delta>0 \; | \; \psi(x) >\psi(x_0), \mbox{for}
\; x \in (x_0-\delta,x_0) \} \neq \emptyset$, since $\psi'(x_0)<0$, so
\[ \delta_0=\sup \mathcal{E} >0. \]
If $\delta_0 <\infty$, then $x_1=x_0-\delta_0$ satisfies
\[  \left\{ \begin{array}{l}
              \psi(x)>\psi(x_0)\ \forall x \in (x_1,x_0)\\
              \psi(x_1)=\psi(x_0).
\end{array} \right. \]
Thus for a.e. $x \in [x_1,x_0)$, we have $\psi''(x) =W(x) \psi(x) \geq W(x)
\psi( x_0) >0$ hence $\psi'(x) < \psi'(x_0)<0$ for all $x \in [x_1,x_0)$,
so we finally get
\[ \psi(x_1) > \psi(x_0). \]
Actually $\psi(x_1)=\psi(x_0)$, hence $\delta_0=+\infty$ and the first
result
follows. Finally, if there is some $x_0 < a$
such that $\psi(x_0) \psi'(x_0) <0$,
then the first result implies
that $|\psi(x)| \geq |\psi(x_0)|>0$, for any $x \leq x_0$. This
is impossible since $\psi \in \Lp{(-\infty,a)}$. $\Box$

\vspace{.1in}
\noindent
We next consider the possibility that the wave function has zeros in the
classically forbidden region.

\begin{lemma}
\label{lm2}
Under the hypotheses of Proposition \ref{prop-negative},
we have $\psi(x) > 0$ for any $x < a$.
\end{lemma}

\vspace{.1in}
\noindent
{\bf Proof.}

\noindent
1. We first show that
$\psi(x) \psi'(x) >0$, for any $x < a$ such that $\psi(x)
\neq 0$.
We assume that $\psi(x)>0$ (the case $\psi(x)<0$ being treated in the same
way) so $\psi(t)>0$ for any $t \in (x-\delta,x)$ for some $\delta>0$ and
$\psi''(t)=W(t) \psi(t) >0$ for a.e. $t$ in $(x-\delta,x)$.
If $\psi'(x)=0$ we have
$\psi'(t) <0$ and also
$\psi(t) \psi'(t) <0$ for each
$t \in (x-\delta,x)$. This is impossible according to Lemma \ref{lm1}.
Hence $\psi' (x)>0$ since $\psi'(x) \geq 0$ by Lemma
\ref{lm1}.

\noindent
2. Next we show that if $\psi(x_0)= 0$,
for some $x_0 < a$, then $\psi'(x_0)=0$.
We assume that $\psi(x_0)= 0$ and $\psi'(x_0)>0$ (the case $\psi'(x_0)<0$
being treated in the same manner).
In this case we can find some $\delta>0$ such that $\psi(x)<0$ and
$\psi'(x)>0$, for any $x \in (x_0-\delta,x_0)$, which is
impossible according to Lemma \ref{lm1}.

\noindent
3. To complete the proof,
we assume that there is a real number $x_0 < a$ such that
$\psi(x_0)=0$. We also have $\psi'(x_0)= 0$ by part 2 and
\[ \sup \{ x < x_0 \; | \; \psi(x) \neq 0 \}=x_0, \]
since $\psi$ would be zero on $(-\infty,a)$ otherwise by Lemma
\ref{lm-UCT}. Thus, we can find some $\delta>0$ such that
$\pm \psi(x)>0$, for all $x \in (x_0-\delta,x_0)$, so $\pm \psi''(x)=W(x)
(\pm \psi(x)) >0$ a.e. in $(x_0-\delta,x_0)$.
This implies
that $\pm \psi'(x)<0$, and, consequently, that
$\psi(x) \psi'(x)<0$, for any $x
\in ( x_0-\delta,x_0)$. This is impossible
according to Lemma \ref{lm1}. $\Box$ \\

To justify the second part of Proposition \ref{prop-negative},
we multiply (\ref{systeme}) by $\psi$,
and integrate over $[x,x_0]$, for some $x_0 <a$ and $x <
x_0$. We obtain
\beq
\label{zz1}
\int_{x}^{x_0} \psi''(u) \psi(u) du= \int_{x}^{x_0} W(u) \psi^2(u) du.
\eeq
Integrating by parts in the left side of (\ref{zz1}),
we get
\beq
\label{zz2}
\psi(x_0) \psi'(x_0) - \psi(x) \psi'(x) - \int_{x}^{x_0} \psi'^2(u) du =
\int_{x}^{x_0} W(u) \psi^2(u) du,
\eeq
so by taking the limit $x \rightarrow -\infty$ in (\ref{zz2}),
we obtain the inequality :
\[ 0 \leq \int_{-\infty}^{x_0} W(u) \psi^2(u) du \leq \psi(x_0) \psi'(x_0)
- \int_{-\infty}^{x_0} \psi'^2(u) du <\infty. \]
Hence, the function $W \psi^2 \in
\Lp[1]{(-\infty,x_0)}$, and the result follows.

\subsection{Pointwise Bounds}

We examine now the behavior of an $\Hp{(-\infty,a)}$-solution to
(\ref{systeme}) for a potential
\beq
\label{WHoneloc}
  W \in \Hploc{(-\infty,a)}.
\eeq
We multiply (\ref{systeme}) by $\psi'(x)$ and integrate over
$[u,t]$, for $u<t<a$ :
\[ \int_u^{t} \psi'(x) \psi''(x) dx  =
   \frac{\psi'^2(t)-\psi'^2(u)}{2} = \int_u^{t} W(x) \psi(x) \psi'(x) dx.
\]
Next, integrating by parts, the right side of this equality gives
\[ \psi'^2(t)-\psi'^2(u) =W(t) \psi^2(t) - W(u) \psi^2(u) - \int_u^{t}
W'(x) \psi^2(x) dx, \]
the above integral being well defined since $W' \in \Lploc{(-\infty,a)}$ and
$\psi$ is bounded in $[u,t]$.
Now taking the limit as $u \rightarrow -\infty$ in the previous equality
leads to
\beq
\label{estimBase}
\psi'^2(t) = W(t) \psi^2(t) - \int_{-\infty}^t W'(u) \psi^2(u) du,\ \forall
t < a ,
\eeq
according to part 2 of Proposition \ref{prop-negative}.
The main result on $L^2$-solutions of the equation (\ref{systeme})
is the following theorem.

\begin{proposition}
\label{prop-estimate}
Let $W$ satisfy conditions
(\ref{positivePot}) and (\ref{WHoneloc}), and be such that
\beq
\label{increasingPot}
W'(x) \leq 0,\ a.e.\ x < a.
\eeq
Then any real $\Hp{(-\infty,a)}$-solution $\psi$ to (\ref{systeme})
satisfies
\[\psi(x_0) \expo{-\int_x^{x_0} \sqrt{S(t)} dt} \leq \psi(x) \leq
  \psi(x_0) \expo{-\int_x^{x_0} \sqrt{W(t)} dt},\ \forall x \leq x_0 \leq
a, \]
where the well-defined function $S(t)$, for $t \leq a$, is given by
\[ S(t) = W(t) - \int_{-\infty}^t W'(u) \expo{-2 \int_u^t \sqrt{W(v)}
dv} du.
\]
\end{proposition}

\vspace{.1in}
\noindent
{\bf Proof.}

\noindent
1. {\bf Upper Bound.} Equality (\ref{estimBase}) combined with
(\ref{increasingPot}) provide $\psi'^2(t) \geq  W(t) \psi^2(t)$, so we have
$\psi'(t) \geq \sqrt{W(t)} \psi(t)$ for any $t < a$, by part 1 of
Proposition \ref{prop-negative}. Next, integrating
over $[x,x_0]$ for $x \leq x_0 < a$, leads to :
\begin{equation}
\label{ub-Wdi}
\psi(x) \leq \psi(x_0) \expo{-\int_x^{x_0} \sqrt{W(t)} dt}.
\end{equation}
Now, the left continuity of $\psi$ at $a$ allows us to extend this
equality at $x_0=a$ by taking the limit in (\ref{ub-Wdi}) as $x_0$
goes to $a$.\\
\noindent
2. {\bf Lower Bound.}
Taking account of (\ref{increasingPot}), it follows readily from
(\ref{estimBase}), together with (\ref{ub-Wdi}), that
$\int_{-\infty}^t W'(u) \expo{-2 \int_u^t \sqrt{W(v)} dv} du<+\infty$ for
any $t <a$. Then, inserting (\ref{ub-Wdi})
written for $u < t <a$
\[ \psi(u) \leq \psi(t) \expo{-\int_{u}^t \sqrt{W(v)} dv}, \]
in (\ref{estimBase}), provides
\[ \psi'^2(t) \leq  S(t) \psi^2(t), \]
for any $t <a$. Thus $\psi'(t) \leq \sqrt{S(t)} \psi(t)$ for all $t < a$,
by part 1 of Proposition \ref{prop-negative}, so we get
\beq
\label{lb-Wdi}
\psi(x) \geq \psi(x_0) \expo{-\int_x^{x_0} \sqrt{S(t)} dt},\ \forall x \leq
x_0 < a,
\eeq
by integrating over $[x,x_0]$.
Taking account of the left continuity of $\psi$ at $a$ we extend this
result at $x_0=a$ by taking the limit in (\ref{lb-Wdi}) as $x_0 \rightarrow
a$. $\Box$ \\

Assumptions (\ref{positivePot}), (\ref{WHoneloc}), and (\ref{increasingPot})
are  essential for
the existence of the upper bound (and also for the lower bound) in
Proposition \ref{prop-estimate}. Nevertheless, the following statement
establishes that $\psi$ remains exponentially increasing in $(-\infty,a)$
for
a bounded from below but non necessarily differentiable or increasing
potential W.

\begin{proposition}
\label{boundedBelow}
If $W \in \Lploc{(-\infty,a)}$ is bounded from below,
\beq
\label{below}
W(x) \geq W_{inf} >0,\ a.e.\ x <a,
\eeq
then any real $\Hp{(-\infty,a)}$-solution of (\ref{systeme}) satisfies :
\[ \psi(x) \leq \psi(x_0) \expo{-W_{inf}^{\demip} (x_0-x)},\ \forall x \leq
x_0 \leq a. \]
\end{proposition}

\vspace{.1in}
\noindent
{\bf Proof.}
We multiply (\ref{systeme}) by $\psi'(u)$ so we get
\[ \psi''(u) \psi'(u) = W(u) \psi(u) \psi'(u) \geq W_{inf} \psi(u)
\psi'(u),\ a.  e.\ u < a, \]
according to (\ref{below}) and Part 1 of Proposition \ref{prop-negative}.
Next we integrate this inequality over $[x,t]$ for $x < t <a$,
\[ \psi'^2(t) - \psi'^2(x) \geq W_{inf} (\psi^2(t) - \psi^2(x)), \]
and take the limit as $x \rightarrow -\infty$ :
\[ \psi'^2(t) \geq W_{inf} \psi^2(t),\ \forall t < a. \]
This leads to $\psi'(t) \geq W_{inf}^{\demip} \psi(t)$ for any $t <a$,
by part 1 of Proposition \ref{prop-negative}. By integrating over
$[x,x_0]$, $x \leq x_0 < a$, we finally obtain
\[ \psi(x) \leq \psi(x_0) \expo{-W_{inf}^{\demip} (x_0-x)}. \]
This result continues to hold for $x_0=a$ since $\psi$ is left continuous
at $a$. $\Box$


\renewcommand{\thechapter}{\arabic{chapter}}
\renewcommand{\thesection}{\thechapter}

\setcounter{chapter}{9} \setcounter{equation}{0}

\section{Appendix 3: Pointwise Bounds for the Eigenfunctions of $h_0(k)$}

We now apply the results of Appendices 1 and 2 to the eigenfunctions
$\varphi_j (.;k)$ of the operator $h_0(k)$.
In appendix 1, we proved each eigenfunction $\varphi_j(.;k)$, $j
\in \N$, of $h_0(k)$, $k \in \R$, is a real $\Hp{\R}$-solution to the
Schr\"odinger equation
\beq
-\varphi_j''(x;k) +V(x;k) \varphi_j(x;k) = \omega_j(k) \varphi_j(x;k),\
\eeq
that is continuously differentiable in $\R$. We now
prove pointwise exponential upper and lower bounds on $\varphi_j(.;k)$
in the classically forbidden region where $W_j(x;k) \equiv V(x;k) -
\omega_j(k) > 0$, based on the general results obtained in Appendix 2.
We prove these bounds under the hypotheses in section 6 on the soft
confining potentials $V_0 \in \Hploc{(-\infty,x_\varepsilon)}$, condition
(H1), and either (H2) or (H2'), which we recall
here:

\begin{tabular}{ll}
& \\
{\bf (H1)} & There is $x_{\varepsilon}$ satisfying (\ref{turningpt2}) for
some
$\varepsilon \in (0,1]$, such that,\\
& $0 \leq V_0(t) \leq (2n + c + 2 / \varepsilon )B \leq V_0(x)$, for all $x
\leq
x_\varepsilon \leq t$.\\
& \\
{\bf (H2)} & $|V_0'(t)| \leq 5 B^{3/2} \slash \sqrt{2 \varepsilon}$
for a.e. $t < x_{\varepsilon}$. \\
& \\
{\bf (H2')} & For any $k \in \Sigma_n$, there is $C_k>0$ such that the
double
inequality,\\
& $-C_k \sqrt{(Bt-k)^2+V_0(t)-(2n+c)B} \leq V_0'(t) + 2B(Bt-k) \leq 0 $,\\
& holds for a.e. $t < x_{\varepsilon}$.\\
& \\
\end{tabular}\\

Let the constant $E$ satisfy $E \leq
(2n+c)B$, $1 < c < 3$, for some $n \in \N$.
We study the behavior of
a real $\Hp{(-\infty,x_{\varepsilon})}$-solution $\psi$ to $(\ref{systeme})$
associated to the perturbed quadratic potential
\[ W(x;k) \equiv (Bx-k)^2 + V_0(x) - E = V(x;k) - E, \]
for some given $k \in \R$.
In the applications, we have $E= \omega_j (k)$, with $k \in \omega_j^{-1}
((- \infty, (2n+c)B])$.

\subsection{Convex-Concave Soft Confining Potentials \\ of Type 1}

We study the soft confining potentials of type 1.
These are distortions of the Sharp Confining Potential
given by
\beq
\label{sharp1}
V_0(x) = \mathcal{V}_0 \chi_{(-\infty, x_\varepsilon)}(x).
\eeq
These confining potentials smooth out the discontinuity at $x=0$ and remain
bounded as $x \rightarrow - \infty$.
The soft confining potentials $V_0 \in \Hploc{(-\infty,x_\varepsilon)}$
are assumed to satisfy conditions (H1)--(H2) of section 6 and restated
above.
With $\mathcal{V}_0 >> (2n+3)B$, these conditions are
obviously satisfied by the Sharp Confining Potential
(\ref{sharp1}), since $V_0'$ is identically zero
in $(-\infty, x_\varepsilon)$ for this model.

We now turn to the general case
and derive the following statement from (H1) and (H2).

\begin{proposition}
\label{prop-sfl}
Let $V_0$ satisfy $(H1)$ and $(H2)$. Then,
for any $k \in \R$ and any $\varepsilon \in (0,1]$, we have
\[ \psi(x_0) \expo{-(1+\varepsilon) \int_{x}^{x_0}
\sqrt{W(t;k)} dt} \leq  \psi(x) \leq \psi(x_0) \expo{-\sqrt{2 B \slash
\varepsilon} (x_0-x)},\
\forall x \leq x_0 \leq x_{\varepsilon}.\]

\end{proposition}

\vspace{.1in}
\noindent
{\bf Proof.}
We fix $k$ in $\R$.\\
\noindent
{\bf Step 1.} The assumption $(H1)$ guarantees
that the effective potential $W(.;k)$ is bounded from below by $2 B \slash
\varepsilon >0$ in
$(-\infty,x_{\varepsilon})$, so
\beq
\label{ub-quadra}
\psi(x) \leq \psi(x_0)
\expo{-\sqrt{2 B \slash \varepsilon} (x_0-x)},\ \forall x \leq x_0 \leq
x_{\varepsilon},
\eeq
by Proposition \ref{boundedBelow}\\
\noindent
{\bf Step 2.} We now prove that the following inequality
\begin{equation}
\label{uk}
\psi'^2(t) \leq U(t;k) \psi^2(t),
\end{equation}
where $U(t;k)=W(t;k)+\varepsilon
B \slash 4 -(\varepsilon B \slash 2)^{\demip} (Bt-k)_-
+5 B \slash 4$ and $(Bt-k)_-=\min ( 0, Bt-k )$,
holds for any $t<x_{\varepsilon}$.
Indeed, by inserting
(\ref{ub-quadra}) written for $u \leq t \leq x_{\varepsilon}$,
\begin{equation}
\label{analogue}
\psi(u) \leq \psi(t) \expo{-\sqrt{2 B \slash \varepsilon}(t-u)},
\end{equation}
in the obvious inequality
\[ \int_{-\infty}^t (Bu-k) \psi^2(u) du \geq
   \int_{-\infty}^{\alpha_k(t)} (Bu-k) \psi^2(u) du, \]
where $\alpha_k(t)=\min(k \slash B,t)$, we get :
\[
\int_{-\infty}^t (Bu-k) \psi^2(u) du \geq
\left( \frac{B \alpha_k(t)-k}{2 \sqrt{2 B \slash \varepsilon}}  -
\frac{\varepsilon}{8} \right)
\expo{-2 \sqrt{2 B \slash \varepsilon} (t- \alpha_k(t))} \psi^2(t). \]
Thus, we deduce from the two following elementary inequalities
$B \alpha_k(t)-k=(Bt-k)_-$
and  $\expo{-2 \sqrt{2 B \slash \varepsilon} (t-\alpha_k(t))} \leq 1$ where
$t < x_{\varepsilon}$, that
\begin{equation}
\label{estimateint}
-\int_{-\infty}^t (Bu-k) \psi^2(u) du
   \leq -\left( \frac{(Bt-k)_-}{2 \sqrt{2B \slash \varepsilon}} -
\frac{\varepsilon}{8} \right) \psi^2(t).
\end{equation}
Since $W(.;k) \in \Hploc{(-\infty,x_{\varepsilon})}$, we have
\[ \psi'^2(t) = W ( t;k) \psi^2(t) - \int_{-\infty}^t (2B(Bu-k) +
V_0'(u)) \psi^2(u) du ,\ a.e.\ t < x_{\varepsilon}, \]
by substituting $W(.;k)$ for $W$ (and also
$2B(Bu-k)+V_0'(u)$ for $W'(u)$) in (\ref{estimBase}),
so (\ref{uk}) follows from this, (\ref{estimateint}) together with
$(H2)$ and (\ref{analogue}).\\
\noindent
{\bf Step 3.} It remains to show that $U(.;k)$ can be made
arbitrarily close from $W(.;k)$ in $(-\infty,x_{\varepsilon})$,
by choosing $\varepsilon$ small enough. To see this we fix $\varepsilon \in
(0,1]$, $t < x_{\varepsilon}$, and deduce from
the basic inequality $-(Bt-k)_- \leq |Bt-k|$ that
\[ U(t;k)-W(t;k)=(\varepsilon+5) B \slash 4 - (\varepsilon B \slash
2)^{\demip}(Bt-k)_-
\leq  R(t;k), \]
where
\[ R(t;k)=\left\{ \begin{array}{lc}
(\varepsilon+5) B \slash 4  + \varepsilon (Bt-k)^2 & \mbox{if}\
(2 \varepsilon B)^{\demip} \left| t - k \slash B \right| \geq 1\\
(\varepsilon +7 ) B \slash 4 & \mbox{if}\
(2 \varepsilon B)^{\demip} \left| t - k \slash B \right| < 1.
\end{array} \right.\]
The condition $0 < \varepsilon  \leq 1$ assures us that
$(\varepsilon +7 ) B \slash 4 \leq 2B$ so
$R(t;k) \leq \varepsilon W(t;k)$, and
\[ U(t;k) \leq (1+\varepsilon) W(t;k),\ \forall t < x_{\varepsilon}. \]
Therefore, it follows from this, (\ref{uk}) and Part 1 of
Proposition \ref{prop-negative} that
\[  \psi(x) \geq \psi(x_0) \expo{-(1+\varepsilon) \int_{x}^{x_0}
\sqrt{W(t;k)} dt},\ \forall x \leq x_0 < x_{\varepsilon}.\]
Moreover, this estimate remains valid at $x_0=x_{\varepsilon}$ since $\psi$
is left
continuous at $x_{\varepsilon}$. $\Box$ \\

\noindent
{\bf Remark:}
Notice that Proposition \ref{prop-sfl} remains valid if we replace
hypothesis (H2) by the weaker hypothesis
\begin{eqnarray*}
& (H2'') & \int_{-\infty}^t | V_0'(u)| \expo{2 \sqrt{2 B \slash \varepsilon}
u} du \leq
\frac{5 B}{4} \expo{2 \sqrt{2 B
\slash \varepsilon} t},\ \forall t < x_{\varepsilon}.
\end{eqnarray*}

\subsection{Parabolic Confining Potential and Soft Confining Potentials of
Type 2}

We first derive the
following general statement from (H1) and (H2').

\begin{proposition}
\label{prop-generale}
Suppose the soft confining potential satisfies (H1) and (H2').
Any given $\varepsilon >0$, we have
\[    \psi(x_0) \expo{-(1+ C_k  \varepsilon \slash B) \int_{x}^{x_0}
\sqrt{W(u;k)}
du} \leq \psi(x) \leq \psi(x_0) \expo{-\int_{x}^{x_0} \sqrt{W(u;k)} du},\
\forall
x \leq x_0 \leq x_{\varepsilon}. \]
\end{proposition}

\vspace{.1in}
\noindent
{\bf Proof.}

\noindent
1. {\bf Upper bound.} The right inequality is a straightforward consequence
of Proposition \ref{prop-estimate} since
$W'(x;k) \leq 0$ for any $x < x_{\varepsilon}$, according to $(H2')$. \\

\noindent
2. {\bf Lower bound.} For any $x < t < x_{\varepsilon}$, we deduce from
$(H2')$ that
\begin{eqnarray*}
\int_{x}^t W'(u;k) \expo{-2 \int_u^t \sqrt{W(v;k)} dv} du & \geq & -C_k
\int_x^{t} \sqrt{W(u;k)} \expo{-2 \int_u^t \sqrt{W(v;k)} dv} du\\
& \geq & -C_k (1-\expo{-2 \int_x^t \sqrt{W(v;k)} dv}) \slash 2,
\end{eqnarray*}
so taking the limit as $x \rightarrow -\infty$ involves
\[ \int_{-\infty}^t W'(u;k) \expo{-2 \int_u^t \sqrt{W(v;k)} dv} du \geq -C_k
\slash 2, \]
since $W(x;k)$ is unbounded as  $x \rightarrow -\infty$. Hence, for any
$t<x_{\varepsilon}$ we have
\begin{eqnarray*}
S(t;k) & = & W(t;k) - \int_{-\infty}^t W'(u) \expo{-2 \int_u^t \sqrt{W(v)}
dv} du \\
        & \leq & W(t;k) + C_k \slash 2\\
        & \leq & \left( 1 + C_k \varepsilon \slash B \right) W(t;k),
\end{eqnarray*}
since $W(t;k) \geq B \slash \varepsilon$.
This shows that $S(.;k) \in \Lploc{(-\infty,x_{\varepsilon})}$ (since
$W(.;k)$ is locally
square integrable on $(-\infty,x_{\varepsilon})$) so
the result follows immediately from Proposition \ref{prop-estimate}. $\Box$
\\

Hypothesis (H2') implies that the soft confining
potential $ V_0$ is, roughly speaking, bounded by two parabolas.
As an example, we show now that the Parabolic Confining Potential,
\beq
\label{para3}
V_0(x)=\mathcal{V}_0 x^2 \chi_{(-\infty,0)}(x),
\eeq
fulfills (H1) and (H2') uniformly for $k \in \Sigma_n =
\cup_{j=0}^n \omega_j^{-1}(\Delta_n)$, where we recall that
$\omega_j(k)$'s are the eigenvalues of $h_0(k)=p_x^2+V(x;k)$.

Any given $\varepsilon \in (0,1]$ and $\mathcal{V}_0>0$, we first
impose $x_{\varepsilon}<0$ satisfies
\beq
\label{condPara}
\mathcal{V}_0 x_{\varepsilon}^2 = (2n+c+ 2 \slash \varepsilon)B,
\eeq
so (H1) is true. We state next with the coming lemma
states that $V_0$ defined by (\ref{para3}) fulfills (H2') for any $k \in
\Sigma_n$.

\begin{lemma}
\label{lm-Wpara}
For any $k \in \Sigma_n$, we have
\[  -\Lambda_n(\varepsilon) \sqrt{W(x;k)} \leq W'(x;k) \leq 0,\ \forall x
\leq  x_{\varepsilon},\]
with $\Lambda_n(\varepsilon) \equiv 2 B_{\mathcal{V}_0} \sqrt{1+ (2n+c)
\varepsilon}$
and $B_{\mathcal{V}_0} \equiv \sqrt{B^2+\mathcal{V}_0}$.
\end{lemma}

\vspace{.1in}
\noindent
{\bf Proof.}

\noindent
1. Let $k$ be negative. Then we have
$V(x;k) \geq \frac{B^2}{B_{\mathcal{V}_0}^2} \tilde{V}(x;k)$ for any $x \in
\R$, where $\tilde{V}(x;k) \equiv (Bx-k)^2 + \mathcal{V}_0 x^2$ is obtained
by
substituting $\mathcal{V}_0 x^2$ for $V_0(x)$ in $V(x;k)$.
Indeed the previous inequality is obvious
when $x < 0$ (since $V(.;k)=\tilde{V}(.;k)$ in $\R_-^*$) and elementary
computations give
\beq
\frac{B_{\mathcal{V}_0}^2}{B^2} V(x;k) 
=\frac{B_{\mathcal{V}_0}^2}{B^2} (B x-k)^2
\tilde{V}(x;k)+ \mathcal{V}_0 \slash B^2 (k^2 - 2 Bk x )   \geq
\tilde{V}(x;k),
\eeq
for $x  \geq 0$. Hence, we have $h_0(k) \geq \tilde{h}_0(k) \equiv p_x^2 +
\frac{B^2}{B_{\mathcal{V}_0}^2}
\tilde{V}(.;k)$ in the operator sense,
so
\beq
\label{vp}
\omega_j(k) \geq (2j+1) B + \frac{\mathcal{V}_0 B^2}{B_{\mathcal{V}_0}^4}
k^2,\ \forall j \in \N,\ \forall k \geq 0,
\eeq
by noticing that
\[ \tilde{h}_0(k) = p_x^2 + (Bx- B^2 \slash B_{\mathcal{V}_0}^2 k )^2 +
\mathcal{V}_0 B^2 \slash B_{\mathcal{V}_0}^4 k^2 \]
is a Landau Hamiltonian with spectrum
$\{ (2j+1) B +  \mathcal{V}_0 B^2 \slash B_{\mathcal{V}_0}^4 k^2,\ j \geq 0
\}$.
Therefore, any $k \leq 0$ such that $\omega_j(k) \leq  (2n+c)B$,
$0 \leq j \leq n$, satisfies
\[ (2j+1) B + \frac{\mathcal{V}_0 B^2}{B_{\mathcal{V}_0}^4} k^2 \leq
(2n+c)B, \]
according to (\ref{vp}), so
\beq
\label{supk}
\inf \left( \Sigma_n \cap \R_- \right)  \geq
-\frac{B_{\mathcal{V}_0}^2}{\sqrt{\mathcal{V}_0} B}\sqrt{(2n+c)B},
\eeq
by recalling that $\dsp{\Sigma_n=\cup_{j=0}^n \omega_j^{-1}(\Delta_n) }$.\\

\noindent
2. Armed with this result we can show now that
\beq
\label{derivPotpos}
W'(x;k) \leq 0,\ \forall x \leq x_{\varepsilon},\ \forall k \in \Sigma_n.
\eeq
Indeed, any given $x \leq x_{\varepsilon} < 0$, the derivative $W'(x;k)=2
B_{\mathcal{V}_0}
(B_{\mathcal{V}_0} x - k B \slash B_{\mathcal{V}_0})$ is obviously negative
for $k  \geq 0$ and we have in addition
\begin{eqnarray*}
W'(x;k) & \leq & 2 B_{\mathcal{V}_0} (B_{\mathcal{V}_0} x_{\varepsilon} - k
B \slash
B_{\mathcal{V}_0})\\
        & \leq & -2 B_{\mathcal{V}_0} (B_{\mathcal{V}_0} \sqrt{(2n+c+2
\slash \varepsilon )B}
\slash \sqrt{\mathcal{V}_0} + k B \slash B_{\mathcal{V}_0}),
\end{eqnarray*}
according to (\ref{condPara}), so
\[  W'(x;k) \leq -\frac{2 B_{\mathcal{V}_0}^2}{\sqrt{\mathcal{V}_0}} \left(
\sqrt{(2n+c+2 \slash \varepsilon) B } - \sqrt{(2n+c)B} \right) < 0, \]
by (\ref{supk}), for any $k < 0$ belonging to $\Sigma_n$.\\

\noindent
3. Let's build now some real number $\lambda >0$, such that
\[ W'(x;k) \geq (-\lambda) \sqrt{W(x;k)},\ \forall x \leq x_{\varepsilon},\
\forall k \in
\Sigma_n.\]
From (\ref{derivPotpos}) this is equivalent to finding $\lambda >0$ such
that $W'(x;k)^2 \leq \lambda^2 W(x;k)$, which leads to
\[ \lambda^2 ( E- k^2 \mathcal{V}_0 \slash B_{\mathcal{V}_0}^2 ) \leq
(\lambda^2 -4 B_{\mathcal{V}_0}^2) (B_{\mathcal{V}_0} x- k B \slash
B_{\mathcal{V}_0})^2,\ \forall k \in \Sigma_n,\]
by noticing that $W(x;k)=(B_{\mathcal{V}_0} x - B \slash B_{\mathcal{V}_0}
k)^2 + \mathcal{V}_0 \slash B_{\mathcal{V}_0}^2 k^2-E$ for $x \leq
x_{\varepsilon}$.
Next, taking account of inequality
\[ (B_{\mathcal{V}_0} x- B \slash B_{\mathcal{V}_0} k)^2 = V(x;k)-
\mathcal{V}_0 \slash B_{\mathcal{V}_0}^2 k^2
\geq (2n+c+ 2 \slash \varepsilon )B - \mathcal{V}_0 \slash
B_{\mathcal{V}_0}^2 k^2, \]
it suffices to find $\lambda$ such that,
\[  \lambda^2 ( E-  \mathcal{V}_0 \slash B_{\mathcal{V}_0}^2 k^2) \leq
(\lambda^2-4 B_{\mathcal{V}_0}^2) \left( (2n+c + 2 \slash \varepsilon )B -
\mathcal{V}_0 \slash
B_{\mathcal{V}_0}^2 k^2 \right), \]
or equivalently
\[ 4 B_{\mathcal{V}_0}^2 \left( (2n+c + 2 \slash \varepsilon)  B  -
\mathcal{V}_0 \slash
B_{\mathcal{ V}_0}^2 k^2 \right)
\leq \lambda^2 \left( (2n+c + 2 \slash \varepsilon)  B  -E \right), \]
for any $k \in \Sigma_n$.
Thus any $\lambda \geq \Lambda_n(\varepsilon)$ is admissible and the result
follows. $\Box$ \\

Having verified conditions (H1)--(H2') for the Parabolic Confining Potential
(\ref{para3}), we summarize the pointwise decay estimates on the
corresponding eigenfunctions following from Proposition
\ref{prop-generale}.

\begin{corollary}
For any $k$ in $\Sigma_n$ and
for any $\varepsilon >0$, the eigenfunctions of $h_0(k)$ with the Parabolic
Confining Potential (\ref{para3}) satisfy the estimates
\[ \psi(x_0) \expo{-(1+ M_{n}(\varepsilon) \varepsilon) \int_{x}^{x_0}
\sqrt{W(u;k) } du}
\leq \psi(x) \leq \psi(x_0) \expo{-\int_{x}^{x_0}
\sqrt{W(u;k)} du},\ \forall x \leq x_0 \leq x_{\varepsilon}, \]
where $M_n(\varepsilon)=2 B_{\mathcal{V}_0} \slash B \sqrt{1 + (2n + c)
\varepsilon}$
and $B_{\mathcal{V}_0} \equiv \sqrt{B^2 + \mathcal{V}_0}$.
\end{corollary}


\end{document}